\documentclass[pre,eqsecnum,preprint,showpacs,preprintnumbers,amsmath,amssymb,amsfonts]{revtex4}

\input epsf

\usepackage{graphicx}

\usepackage{bm}

\newcommand{\beq}{\begin{equation}}
\newcommand{\eeq}{\end{equation}}
\newcommand{\beqs}{\begin{eqnarray}}
\newcommand{\eeqs}{\end{eqnarray}}

\newtheorem{theo}{Theorem}[section]

\newtheorem{lemma}[theo]{Lemma}
\newtheorem{defi}{Definition}[section]

\newtheorem{propo}{Proposition}[section]

\begin{document}

\pagestyle{myheadings}
\markright{Dimer coverings on the Sierpinski gasket}

\title{Dimer coverings on the Sierpinski gasket with possible vacancies on the outmost vertices}

\author{Shu-Chiuan Chang$^{a,b}$} 
\email{scchang@mail.ncku.edu.tw} 

\affiliation{(a) \ Department of Physics \\
National Cheng Kung University \\
Tainan 70101, Taiwan} 

\affiliation{(b) \ Physics Division \\
National Center for Theoretical Science \\
National Taiwan University \\
Taipei 10617, Taiwan} 

\author{Lung-Chi Chen$^{c}$} 
\email{lcchen@math.fju.edu.tw}
\altaffiliation{This paper is written during the author's visit to PIMS, University of British Columbia. The author thanks the institute for the hospitality.}

\affiliation{(c) \ Department of Mathematics \\
Fu Jen Catholic University \\
Taipei 24205, Taiwan }

\date{\today}

\begin{abstract}

We present the number of dimers $N_d(n)$ on the Sierpinski gasket $SG_d(n)$ at stage $n$ with dimension $d$ equal to two, three, four or five, where one of the outmost vertices is not covered when the number of vertices $v(n)$ is an odd number. The entropy of absorption of diatomic molecules per site, defined as $S_{SG_d}=\lim_{n \to \infty} \ln N_d(n)/v(n)$, is calculated to be $\ln(2)/3$ exactly for $SG_2(n)$. The numbers of dimers on the generalized Sierpinski gasket $SG_{d,b}(n)$ with $d=2$ and $b=3,4,5$ are also obtained exactly. Their entropies are equal to $\ln(6)/7$, $\ln(28)/12$, $\ln(200)/18$, respectively. The upper and lower bounds for the entropy are derived in terms of the results at a certain stage for $SG_d(n)$ with $d=3,4,5$. As the difference between these bounds converges quickly to zero as the calculated stage increases, the numerical value of $S_{SG_d}$ with $d=3,4,5$ can be evaluated with more than a hundred significant figures accurate. 


\end{abstract}

\maketitle

\section{Introduction}
\label{sectionI}

The enumeration of close-packed dimers $N(G)$ on a graph $G$ was first considered by Fowler and Rushbrooke in enumerating the absorption of diatomic molecules on a surface \cite{fowler}. The dimer coverings of a graph is a classical model in statistical physics and is called perfect matchings in mathematical literature. The dimer model on the square lattice was solved exactly by Kasteleyn \cite{kasteleyn61} and Temperley and Fisher \cite{temperley61, fisher61}. The model is equivalent to various other statistical mechanical problems. For example, the zero-field partition function of Ising model on a planar lattice can be formulated as a dimer model on an associated planar lattice \cite{kasteleyn63, fisher66}. It is also well known that there is a bijection between close-packed dimer coverings and spanning tree configurations on two related planar lattices \cite{temperley74}. A recent review on the enumeration of close-packed dimers on two-dimensional regular lattices is summarized in Ref. \cite{wu06a}. It is of interest to consider dimer coverings on self-similar fractal lattices which have scaling invariance rather than translational invariance. Fractals are geometric structures of noninteger Hausdorff dimension realized by repeated construction of an elementary shape on progressively smaller length scales \cite{mandelbrot,Falconer}. A well-known example of fractal is the Sierpinski gasket which has been extensively studied in several contexts \cite{Gefen80,Gefen81,Rammal,Alexander,Domany,Gefen8384,Guyer,Kusuoka,Dhar97,Daerden,Dhar05}. A dimer coverings will leave at least one vertex uncovered when the total number of vertices is an odd number, e.g., the rectangular lattice with both length and width odd \cite{tzeng,wu06b}. The vacancies that are not covered by any dimers can be considered as occupied by monomers. We allow such possible vacancies occur on the outmost vertices of the Sierpinski gasket. We shall derive rigorously the numbers of dimer coverings on the two-dimensional Sierpinski gasket and its generalization, and obtain upper and lower bounds for the entropy on the Sierpinski gasket with dimension equal to three, four or five.

\section{Preliminaries}
\label{sectionII}

We first recall some relevant definitions in this section. A connected graph (without loops) $G=(V,E)$ is defined by its vertex (site) and edge (bond) sets $V$ and $E$ \cite{bbook,fh}.  Let $v(G)=|V|$ be the number of vertices and $e(G)=|E|$ the number of edges in $G$.  The degree or coordination number $k_i$ of a vertex $v_i \in V$ is the number of edges attached to it.  A $k$-regular graph is a graph with the property that each of its vertices has the same degree $k$. In general, one can associate a dimer (monomer) weight to each dimer (monomer) (see, for example \cite{tzeng}). For simplicity, all dimer (monomer) weights are set to one throughout this paper. 

When the size of the graph increases as $v(G) \to \infty$, the number of dimer coverings $N(G)$ grows exponentially in $v(G)$. The entropy of absorption of diatomic molecules per site is given by
\beq
S_G = \lim_{v(G) \to \infty} \frac{\ln N(G)}{v(G)} \ ,
\label{sdef}
\eeq
where $G$, when used as a subscript in this manner, implicitly refers to
the thermodynamic limit. The dimer coverings considered here may not be close-packed dimers since there may be vacancies on the outmost vertices as mentioned above. Notice that we define the entropy per site rather than entropy per dimer. They differ by a factor of two in the thermodynamic limit regardless the presence of vacancies on the outmost vertices.

\begin{figure}[htbp]
\unitlength 0.9mm \hspace*{3mm}
\begin{picture}(108,40)
\put(0,0){\line(1,0){6}}
\put(0,0){\line(3,5){3}}
\put(6,0){\line(-3,5){3}}
\put(3,-4){\makebox(0,0){$SG_2(0)$}}
\put(12,0){\line(1,0){12}}
\put(12,0){\line(3,5){6}}
\put(24,0){\line(-3,5){6}}
\put(15,5){\line(1,0){6}}
\put(18,0){\line(3,5){3}}
\put(18,0){\line(-3,5){3}}
\put(18,-4){\makebox(0,0){$SG_2(1)$}}
\put(30,0){\line(1,0){24}}
\put(30,0){\line(3,5){12}}
\put(54,0){\line(-3,5){12}}
\put(36,10){\line(1,0){12}}
\put(42,0){\line(3,5){6}}
\put(42,0){\line(-3,5){6}}
\multiput(33,5)(12,0){2}{\line(1,0){6}}
\multiput(36,0)(12,0){2}{\line(3,5){3}}
\multiput(36,0)(12,0){2}{\line(-3,5){3}}
\put(39,15){\line(1,0){6}}
\put(42,10){\line(3,5){3}}
\put(42,10){\line(-3,5){3}}
\put(42,-4){\makebox(0,0){$SG_2(2)$}}
\put(60,0){\line(1,0){48}}
\put(72,20){\line(1,0){24}}
\put(60,0){\line(3,5){24}}
\put(84,0){\line(3,5){12}}
\put(84,0){\line(-3,5){12}}
\put(108,0){\line(-3,5){24}}
\put(66,10){\line(1,0){12}}
\put(90,10){\line(1,0){12}}
\put(78,30){\line(1,0){12}}
\put(72,0){\line(3,5){6}}
\put(96,0){\line(3,5){6}}
\put(84,20){\line(3,5){6}}
\put(72,0){\line(-3,5){6}}
\put(96,0){\line(-3,5){6}}
\put(84,20){\line(-3,5){6}}
\multiput(63,5)(12,0){4}{\line(1,0){6}}
\multiput(66,0)(12,0){4}{\line(3,5){3}}
\multiput(66,0)(12,0){4}{\line(-3,5){3}}
\multiput(69,15)(24,0){2}{\line(1,0){6}}
\multiput(72,10)(24,0){2}{\line(3,5){3}}
\multiput(72,10)(24,0){2}{\line(-3,5){3}}
\multiput(75,25)(12,0){2}{\line(1,0){6}}
\multiput(78,20)(12,0){2}{\line(3,5){3}}
\multiput(78,20)(12,0){2}{\line(-3,5){3}}
\put(81,35){\line(1,0){6}}
\put(84,30){\line(3,5){3}}
\put(84,30){\line(-3,5){3}}
\put(84,-4){\makebox(0,0){$SG_2(3)$}}
\end{picture}

\vspace*{5mm}
\caption{\footnotesize{The first four stages $n=0,1,2,3$ of the two-dimensional Sierpinski gasket $SG_2(n)$.}} 
\label{sgfig}
\end{figure}
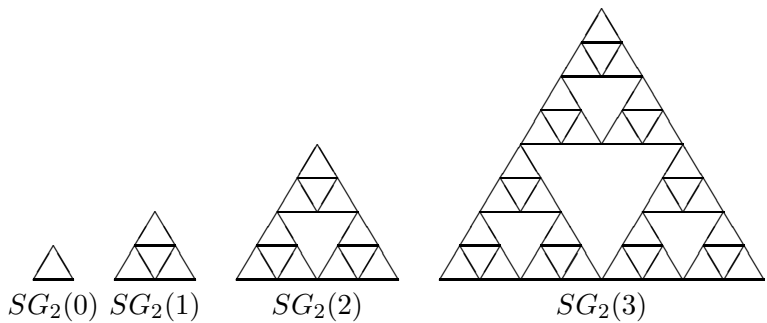

The construction of the two-dimensional Sierpinski gasket $SG_2(n)$ at stage $n$ is shown in Fig. \ref{sgfig}. At stage $n=0$, it is an equilateral triangle; while stage $n+1$ is obtained by the juxtaposition of three $n$-stage structures. In general, the Sierpinski gaskets $SG_d$ can be built in any Euclidean dimension $d$ with fractal dimensionality $D=\ln(d+1)/\ln2$ \cite{Gefen81}. For the Sierpinski gasket $SG_d(n)$, the numbers of edges and vertices are given by 
\beq
e(SG_d(n)) = {d+1 \choose 2} (d+1)^n = \frac{d}{2} (d+1)^{n+1} \ ,
\label{e}
\eeq
\beq
v(SG_d(n)) = \frac{d+1}{2} [(d+1)^n+1] \ .
\label{v}
\eeq
Except the $(d+1)$ outmost vertices which have degree $d$, all other vertices of $SG_d(n)$ have degree $2d$. In the large $n$ limit, $SG_d$ is $2d$-regular. 

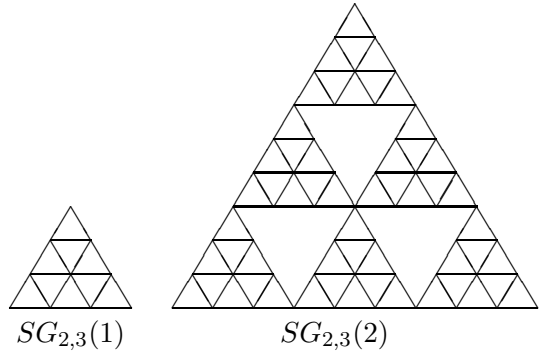
\begin{figure}[htbp]
\unitlength 0.9mm \hspace*{3mm}
\begin{picture}(78,45)
\put(0,0){\line(1,0){18}}
\put(3,5){\line(1,0){12}}
\put(6,10){\line(1,0){6}}
\put(0,0){\line(3,5){9}}
\put(6,0){\line(3,5){6}}
\put(12,0){\line(3,5){3}}
\put(18,0){\line(-3,5){9}}
\put(12,0){\line(-3,5){6}}
\put(6,0){\line(-3,5){3}}
\put(9,-4){\makebox(0,0){$SG_{2,3}(1)$}}
\put(24,0){\line(1,0){54}}
\put(33,15){\line(1,0){36}}
\put(42,30){\line(1,0){18}}
\put(24,0){\line(3,5){27}}
\put(42,0){\line(3,5){18}}
\put(60,0){\line(3,5){9}}
\put(78,0){\line(-3,5){27}}
\put(60,0){\line(-3,5){18}}
\put(42,0){\line(-3,5){9}}
\multiput(27,5)(18,0){3}{\line(1,0){12}}
\multiput(30,10)(18,0){3}{\line(1,0){6}}
\multiput(30,0)(18,0){3}{\line(3,5){6}}
\multiput(36,0)(18,0){3}{\line(3,5){3}}
\multiput(36,0)(18,0){3}{\line(-3,5){6}}
\multiput(30,0)(18,0){3}{\line(-3,5){3}}
\multiput(36,20)(18,0){2}{\line(1,0){12}}
\multiput(39,25)(18,0){2}{\line(1,0){6}}
\multiput(39,15)(18,0){2}{\line(3,5){6}}
\multiput(45,15)(18,0){2}{\line(3,5){3}}
\multiput(45,15)(18,0){2}{\line(-3,5){6}}
\multiput(39,15)(18,0){2}{\line(-3,5){3}}
\put(45,35){\line(1,0){12}}
\put(48,40){\line(1,0){6}}
\put(48,30){\line(3,5){6}}
\put(54,30){\line(3,5){3}}
\put(54,30){\line(-3,5){6}}
\put(48,30){\line(-3,5){3}}
\put(48,-4){\makebox(0,0){$SG_{2,3}(2)$}}
\end{picture}

\vspace*{5mm}
\caption{\footnotesize{The generalized two-dimensional Sierpinski gasket $SG_{2,b}(n)$ with $b=3$ at stage $n=1, 2$.}} 
\label{sgbfig1}
\end{figure}

\begin{figure}[htbp]
\unitlength 0.9mm \hspace*{3mm}
\begin{picture}(60,45)
\put(0,0){\line(1,0){24}}
\put(3,5){\line(1,0){18}}
\put(6,10){\line(1,0){12}}
\put(9,15){\line(1,0){6}}
\put(0,0){\line(3,5){12}}
\put(6,0){\line(3,5){9}}
\put(12,0){\line(3,5){6}}
\put(18,0){\line(3,5){3}}
\put(24,0){\line(-3,5){12}}
\put(18,0){\line(-3,5){9}}
\put(12,0){\line(-3,5){6}}
\put(6,0){\line(-3,5){3}}
\put(12,-4){\makebox(0,0){$SG_{2,4}(1)$}}
\put(30,0){\line(1,0){30}}
\put(33,5){\line(1,0){24}}
\put(36,10){\line(1,0){18}}
\put(39,15){\line(1,0){12}}
\put(42,20){\line(1,0){6}}
\put(30,0){\line(3,5){15}}
\put(36,0){\line(3,5){12}}
\put(42,0){\line(3,5){9}}
\put(48,0){\line(3,5){6}}
\put(54,0){\line(3,5){3}}
\put(60,0){\line(-3,5){15}}
\put(54,0){\line(-3,5){12}}
\put(48,0){\line(-3,5){9}}
\put(42,0){\line(-3,5){6}}
\put(36,0){\line(-3,5){3}}
\put(42,-4){\makebox(0,0){$SG_{2,5}(1)$}}
\end{picture}

\vspace*{5mm}
\caption{\footnotesize{The generalized two-dimensional Sierpinski gasket $SG_{2,b}(n)$ with $b=4,5$ at stage $n=1$.}} 
\label{sgbfig2}
\end{figure}
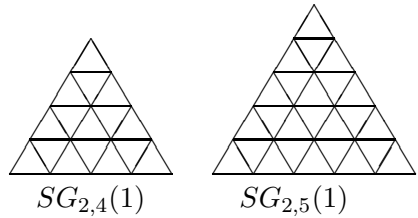

The Sierpinski gasket can be generalized, denoted as $SG_{d,b}(n)$, by introducing the side length $b$ which is an integer larger or equal to two \cite{Hilfer}. The generalized Sierpinski gasket at stage $n+1$ is constructed with $b$ layers of stage $n$ hypertetrahedrons. The two-dimensional $SG_{2,b}(n)$ with $b=3$ at stage $n=1, 2$ are illustrated in Fig. \ref{sgbfig1}, and those with $b=4,5$ at stage $n=1$ in Fig. \ref{sgbfig2}. The ordinary Sierpinski gasket $SG_d(n)$ corresponds to the $b=2$ case, where the index $b$ is neglected for simplicity. The Hausdorff dimension for $SG_{d,b}$ is given by $D=\ln {b+d-1 \choose d} / \ln b$ \cite{Hilfer}. For the two-dimensional Sierpinski gasket $SG_{2,b}(n)$ that will be considered here, the numbers of edges and vertices are given by 
\beq
e(SG_{2,b}(n)) = 3 \Big [\frac{b(b+1)}{2} \Big ]^n \ ,
\label{be}
\eeq
\beq
v(SG_{2,b}(n)) = \frac{b+4}{b+2} \Big [\frac{b(b+1)}{2} \Big ]^n + \frac{2(b+1)}{b+2} \ .
\label{bv}
\eeq
Notice that $SG_{d,b}$ is not $k$-regular even in the thermodynamic limit.

\section{The number of dimer coverings on $SG_{2,b}(n)$ with $b=2,3,4,5$}
\label{sectionIII}

\begin{figure}[htbp]
\unitlength 1.8mm 
\begin{picture}(42,5)
\put(0,0){\line(1,0){6}}
\put(0,0){\line(3,5){3}}
\put(6,0){\line(-3,5){3}}
\multiput(1,0.5)(4,0){2}{\circle{1}}
\put(3,4){\circle{1}}
\put(3,-2){\makebox(0,0){$f_{2,b}(n)$}}
\put(12,0){\line(1,0){6}}
\put(12,0){\line(3,5){3}}
\put(18,0){\line(-3,5){3}}
\put(13,0.5){\circle{1}}
\put(17,0.5){\circle{1}}
\put(15,4){\circle*{1}}
\put(15,-2){\makebox(0,0){$g_{2,b}(n)$}}
\put(24,0){\line(1,0){6}}
\put(24,0){\line(3,5){3}}
\put(30,0){\line(-3,5){3}}
\put(25,0.5){\circle*{1}}
\put(29,0.5){\circle*{1}}
\put(27,4){\circle{1}}
\put(27,-2){\makebox(0,0){$h_{2,b}(n)$}}
\put(36,0){\line(1,0){6}}
\put(36,0){\line(3,5){3}}
\put(42,0){\line(-3,5){3}}
\multiput(37,0.5)(4,0){2}{\circle*{1}}
\put(39,4){\circle*{1}}
\put(39,-2){\makebox(0,0){$t_{2,b}(n)$}}
\end{picture}

\vspace*{5mm}
\caption{\footnotesize{Illustration for the configurations $f_{2,b}(n)$, $g_{2,b}(n)$, $h_{2,b}(n)$, and $t_{2,b}(n)$. Only the three outmost vertices are shown explicitly, where each open circle is vacant and each solid circle is occupied by a dimer.}} 
\label{fghtfig}
\end{figure}
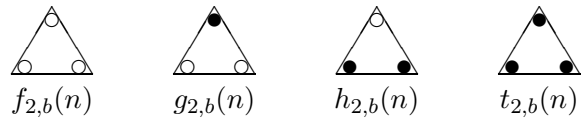

In this section we derive rigorously the numbers of dimer coverings on the two-dimensional Sierpinski gasket $SG_2(n)$, equivalently $SG_{2,2}(n)$, and the generalized $SG_{2,b}(n)$ with $b=3,4,5$. Let us start with the definitions of the quantities to be used. They are illustrated in Fig. \ref{fghtfig}, where only the outmost vertices of $SG_{2,b}(n)$ are shown.

\bigskip

\begin{defi} \label{defisg2} Consider the generalized two-dimensional Sierpinski gasket $SG_{2,b}(n)$ at stage $n$. (i) Define $f_{2,b}(n)$ as the number of dimer coverings such that the three outmost vertices are vacant. (ii) Define $g_{2,b}(n)$ as the numbers of dimer coverings such that one certain outmost vertex, say the topmost vertex as illustrated in Fig. \ref{fghtfig}, is occupied by a dimer while the other two outmost vertices are vacant. (iii) Define $h_{2,b}(n)$ as the numbers of dimer coverings such that one certain outmost vertex, say the topmost vertex as illustrated in Fig. \ref{fghtfig}, is vacant while the other two outmost vertices are occupied by dimers. (iv) Define $t_{2,b}(n)$ as the number of dimer coverings such that all three outmost vertices are occupied by dimers. 
\end{defi}

\bigskip

\subsection{$SG_2(n)$}

For the ordinary two-dimensional Sierpinski gasket, we use the notations $f_2(n)$, $g_2(n)$, $h_2(n)$, and $t_2(n)$ for simplicity. Because of rotational symmetry, there are three possible $g_2(n)$ and three possible $h_2(n)$ for non-negative integer $n$. The initial values at stage zero are $f_2(0)=1$, $g_2(0)=0$, $h_2(0)=1$, $t_2(0)=0$. The values at stage one are $f_2(1)=0$, $g_2(1)=2$, $h_2(1)=0$, $t_2(1)=2$. The value zero indicates that no such configurations are allowed. By Eq. (\ref{v}), we have 
\beqs
v(SG_2(n)) & = & \frac32 (3^n+1) = 3^n + 2 + \frac12 (3^n-1) \cr\cr
& = & 3^n + 2 + \sum_{j=1}^n {n \choose j} 2^{j-1} 
= 3^n + 2 + n + \sum_{j=2}^n {n \choose j} 2^{j-1} \ ,
\eeqs
where the Binomial expansion is used for $3^n=(2+1)^n$, such that the number of vertices for $SG_2(n)$ is odd for even $n$ and even for odd $n$. Therefore, $f(n)$, $h(n)$ are always zero for odd $n$ and $g(n)$, $t(n)$ are always zero for even $n$. Let us denote odd $n$ as $2m+1$ and even $n$ as $2m$ with non-negative integer $m$ in the following discussion for $SG_2(n)$. These quantities satisfy simple recursion relations.

\bigskip

\begin{lemma} \label{lemmasg2r} For any odd $n=2m+1>0$,
\beq
f_2(2m+2) = 2 g_2^3(2m+1) \ , 
\label{feq}
\eeq
\beq
h_2(2m+2) = 2 g_2^2(2m+1) t_2(2m+1) \ . 
\label{heq}
\eeq
For any even $n=2m\ge 0$,
\beq
g_2(2m+1) = 2 f_2(2m) h_2^2(2m) \ , 
\label{geq}
\eeq
\beq
t_2(2m+1) = 2 h_2^3(2m)  \ . 
\label{teq}
\eeq
\end{lemma}

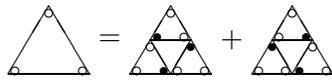
\begin{figure}[htbp]
\unitlength 0.9mm 
\begin{picture}(48,12)
\put(0,0){\line(1,0){12}}
\put(0,0){\line(3,5){6}}
\put(12,0){\line(-3,5){6}}
\multiput(1,0.5)(10,0){2}{\circle{1}}
\put(6,9){\circle{1}}
\put(15,5){\makebox(0,0){$=$}}
\put(18,0){\line(1,0){12}}
\put(18,0){\line(3,5){6}}
\put(24,0){\line(3,5){3}}
\put(24,0){\line(-3,5){3}}
\put(21,5){\line(1,0){6}}
\put(30,0){\line(-3,5){6}}
\put(19,0.5){\circle{1}}
\put(21,4){\circle{1}}
\put(23,0.5){\circle*{1}}
\put(22,5.5){\circle*{1}}
\put(24,9){\circle{1}}
\put(26,5.5){\circle{1}}
\put(25,0.5){\circle{1}}
\put(27,4){\circle*{1}}
\put(29,0.5){\circle{1}}
\put(33,5){\makebox(0,0){$+$}}
\put(36,0){\line(1,0){12}}
\put(36,0){\line(3,5){6}}
\put(42,0){\line(3,5){3}}
\put(42,0){\line(-3,5){3}}
\put(39,5){\line(1,0){6}}
\put(48,0){\line(-3,5){6}}
\put(37,0.5){\circle{1}}
\put(39,4){\circle*{1}}
\put(41,0.5){\circle{1}}
\put(40,5.5){\circle{1}}
\put(42,9){\circle{1}}
\put(44,5.5){\circle*{1}}
\put(43,0.5){\circle*{1}}
\put(45,4){\circle{1}}
\put(47,0.5){\circle{1}}
\end{picture}

\caption{\footnotesize{Illustration for the expression of $f_2(2m+2)$.}} 
\label{ffig}
\end{figure}

{\sl Proof} \quad 
The Sierpinski gasket $SG_2(n+1)$ is composed of three $SG_2(n)$ with three pairs of vertices identified. For each pair of identified vertices, either one of them is originally occupied by a dimer while the other one is vacant. The number $f_2(2m+2)$ for $SG_2(2m+2)$ consists of two configurations where all three of the $SG_2(2m+1)$ are in the $g_2(2m+1)$ status as illustrated in Fig. \ref{ffig}, such that Eq. (\ref{feq}) is verified. 

\bigskip

\bigskip

\begin{figure}[htbp]
\unitlength 0.9mm 
\begin{picture}(48,12)
\put(0,0){\line(1,0){12}}
\put(0,0){\line(3,5){6}}
\put(12,0){\line(-3,5){6}}
\multiput(1,0.5)(10,0){2}{\circle*{1}}
\put(6,9){\circle{1}}
\put(15,5){\makebox(0,0){$=$}}
\put(18,0){\line(1,0){12}}
\put(18,0){\line(3,5){6}}
\put(24,0){\line(3,5){3}}
\put(24,0){\line(-3,5){3}}
\put(21,5){\line(1,0){6}}
\put(30,0){\line(-3,5){6}}
\put(19,0.5){\circle*{1}}
\put(21,4){\circle{1}}
\put(23,0.5){\circle{1}}
\put(22,5.5){\circle*{1}}
\put(24,9){\circle{1}}
\put(26,5.5){\circle{1}}
\put(25,0.5){\circle*{1}}
\put(27,4){\circle*{1}}
\put(29,0.5){\circle*{1}}
\put(33,5){\makebox(0,0){$+$}}
\put(36,0){\line(1,0){12}}
\put(36,0){\line(3,5){6}}
\put(42,0){\line(3,5){3}}
\put(42,0){\line(-3,5){3}}
\put(39,5){\line(1,0){6}}
\put(48,0){\line(-3,5){6}}
\put(37,0.5){\circle*{1}}
\put(39,4){\circle*{1}}
\put(41,0.5){\circle*{1}}
\put(40,5.5){\circle{1}}
\put(42,9){\circle{1}}
\put(44,5.5){\circle*{1}}
\put(43,0.5){\circle{1}}
\put(45,4){\circle{1}}
\put(47,0.5){\circle*{1}}
\end{picture}

\caption{\footnotesize{Illustration for the expression of $h_2(2m+2)$.}} 
\label{hfig}
\end{figure}
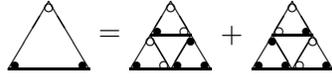

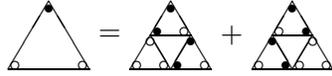
\begin{figure}[htbp]
\unitlength 0.9mm 
\begin{picture}(48,12)
\put(0,0){\line(1,0){12}}
\put(0,0){\line(3,5){6}}
\put(12,0){\line(-3,5){6}}
\multiput(1,0.5)(10,0){2}{\circle{1}}
\put(6,9){\circle*{1}}
\put(15,5){\makebox(0,0){$=$}}
\put(18,0){\line(1,0){12}}
\put(18,0){\line(3,5){6}}
\put(24,0){\line(3,5){3}}
\put(24,0){\line(-3,5){3}}
\put(21,5){\line(1,0){6}}
\put(30,0){\line(-3,5){6}}
\put(19,0.5){\circle{1}}
\put(21,4){\circle{1}}
\put(23,0.5){\circle{1}}
\put(22,5.5){\circle*{1}}
\put(24,9){\circle*{1}}
\put(26,5.5){\circle{1}}
\put(25,0.5){\circle*{1}}
\put(27,4){\circle*{1}}
\put(29,0.5){\circle{1}}
\put(33,5){\makebox(0,0){$+$}}
\put(36,0){\line(1,0){12}}
\put(36,0){\line(3,5){6}}
\put(42,0){\line(3,5){3}}
\put(42,0){\line(-3,5){3}}
\put(39,5){\line(1,0){6}}
\put(48,0){\line(-3,5){6}}
\put(37,0.5){\circle{1}}
\put(39,4){\circle*{1}}
\put(41,0.5){\circle*{1}}
\put(40,5.5){\circle{1}}
\put(42,9){\circle*{1}}
\put(44,5.5){\circle*{1}}
\put(43,0.5){\circle{1}}
\put(45,4){\circle{1}}
\put(47,0.5){\circle{1}}
\end{picture}

\caption{\footnotesize{Illustration for the expression of $g_2(2m+1)$.}} 
\label{gfig}
\end{figure}

\begin{figure}[htbp]
\unitlength 0.9mm 
\begin{picture}(48,12)
\put(0,0){\line(1,0){12}}
\put(0,0){\line(3,5){6}}
\put(12,0){\line(-3,5){6}}
\multiput(1,0.5)(10,0){2}{\circle*{1}}
\put(6,9){\circle*{1}}
\put(15,5){\makebox(0,0){$=$}}
\put(18,0){\line(1,0){12}}
\put(18,0){\line(3,5){6}}
\put(24,0){\line(3,5){3}}
\put(24,0){\line(-3,5){3}}
\put(21,5){\line(1,0){6}}
\put(30,0){\line(-3,5){6}}
\put(19,0.5){\circle*{1}}
\put(21,4){\circle{1}}
\put(23,0.5){\circle*{1}}
\put(22,5.5){\circle*{1}}
\put(24,9){\circle*{1}}
\put(26,5.5){\circle{1}}
\put(25,0.5){\circle{1}}
\put(27,4){\circle*{1}}
\put(29,0.5){\circle*{1}}
\put(33,5){\makebox(0,0){$+$}}
\put(36,0){\line(1,0){12}}
\put(36,0){\line(3,5){6}}
\put(42,0){\line(3,5){3}}
\put(42,0){\line(-3,5){3}}
\put(39,5){\line(1,0){6}}
\put(48,0){\line(-3,5){6}}
\put(37,0.5){\circle*{1}}
\put(39,4){\circle*{1}}
\put(41,0.5){\circle{1}}
\put(40,5.5){\circle{1}}
\put(42,9){\circle*{1}}
\put(44,5.5){\circle*{1}}
\put(43,0.5){\circle*{1}}
\put(45,4){\circle{1}}
\put(47,0.5){\circle*{1}}
\end{picture}

\caption{\footnotesize{Illustration for the expression of $t_2(2m+1)$.}} 
\label{tfig}
\end{figure}
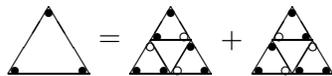

Similarly, $h_2(2m+2)$ and $g_2(2m+1)$, $t_2(2m+1)$ can be obtained with appropriate configurations of its three constituting blocks as illustrated in Figs. \ref{hfig}, \ref{gfig} and \ref{tfig} to verify Eqs. (\ref{heq}), (\ref{geq}) and (\ref{teq}), respectively. \ $\Box$

\bigskip

\bigskip

\bigskip

\bigskip

It is elementary to solve $f_2(n)$, $g_2(n)$, $h_2(n)$, $t_2(n)$ in order to obtain the entropy for $SG_2$.

\bigskip

\begin{theo} \label{theosg2} For the two-dimensional Sierpinski gasket $SG_2(n)$ at stage $n=2m$ or $n=2m+1$,
\beq
\begin{cases}
f_2(2m) = h_2(2m) = 2^{\gamma_2 (2m)} \cr
        f_2(2m+1) = h_2(2m+1) = 0 
\end{cases} \ ,
\label{fh2}
\eeq
\beq
\begin{cases}
g_2(2m) = t_2(2m) = 0 \cr
        g_2(2m+1) = t_2(2m+1) = 2^{\gamma_2 (2m+1)} 
\end{cases} \ ,
\label{gt2}
\eeq
where the exponent is
\beq
\gamma_2 (n) = \frac12 (3^n-1) \ .
\label{alpha2}
\eeq
Define the number of dimer coverings $N(SG_2(n))$ in Eq. (\ref{sdef}) equal to $f_2(n=2m)$ and equal to $g_2(n=2m+1)$ for even and odd $n$, respectively. With $v(SG_2(n))=\frac32 (3^n+1)$, the entropy is given by 
\beq
S_{SG_2} = \frac13 \ln 2 \simeq 0.23104906018...
\label{Ssg2}
\eeq
\end{theo}

\subsection{$SG_{2,3}(n)$}

For the generalized two-dimensional Sierpinski gasket $SG_{2,b}(n)$ with $b=3$, we have
\beq
v(SG_{2,3}(n)) = \frac{7(6)^n+8}{5} = 6^n + 2 + \frac25 (6^n-1) 
= 6^n + 2 + 2\sum_{j=1}^n {n \choose j} 5^{j-1} 
\eeq
by Eq. (\ref{bv}), such that the number of vertices is equal to three for $n=0$ and becomes even for all positive integer $n$. Therefore, $f_{2,3}(n)$ and $h_{2,3}(n)$ are always zero for positive integer $n$, while the initial values remain $f_{2,3}(0)=1$, $g_{2,3}(0)=0$, $h_{2,3}(0)=1$ and $t_{2,3}(0)=0$. $g_{2,3}(n)$ and $t_{2,3}(n)$ satisfy recursion relations.

\bigskip

\begin{lemma} \label{lemmasg23r} For any positive integer $n$,
\beq
g_{2,3}(n+1) = 6 g_{2,3}^5(n) t_{2,3}(n) \ , 
\label{g23eq}
\eeq
\beq
t_{2,3}(n+1) = 6 g_{2,3}^4(n) t_{2,3}^2(n) \ ,
\label{t23eq}
\eeq
and for $n=0$,
\beq
g_{2,3}(1) = 6 f_{2,3}^2(0) h_{2,3}^4(0) = 6 \ , 
\label{g231eq}
\eeq
\beq
t_{2,3}(1) = 6 f_{2,3}(0) h_{2,3}^5(0) = 6 \ . 
\label{t231eq}
\eeq
\end{lemma}

\begin{figure}[htbp]
\unitlength 0.9mm 
\begin{picture}(118,18)
\put(0,0){\line(1,0){18}}
\put(0,0){\line(3,5){9}}
\put(18,0){\line(-3,5){9}}
\multiput(1,0.5)(16,0){2}{\circle{1}}
\put(9,14){\circle*{1}}
\put(21,5){\makebox(0,0){$=$}}
\put(24,0){\line(1,0){18}}
\put(27,5){\line(1,0){12}}
\put(30,10){\line(1,0){6}}
\put(24,0){\line(3,5){9}}
\put(30,0){\line(3,5){6}}
\put(36,0){\line(3,5){3}}
\put(30,0){\line(-3,5){3}}
\put(36,0){\line(-3,5){6}}
\put(42,0){\line(-3,5){9}}
\put(25,0.5){\circle{1}}
\put(27,4){\circle{1}}
\put(29,0.5){\circle*{1}}
\put(31,0.5){\circle{1}}
\put(33,4){\circle*{1}}
\put(35,0.5){\circle{1}}
\put(37,0.5){\circle*{1}}
\put(39,4){\circle{1}}
\put(41,0.5){\circle{1}}
\put(28,5.5){\circle*{1}}
\put(30,9){\circle{1}}
\put(32,5.5){\circle{1}}
\put(34,5.5){\circle{1}}
\put(36,9){\circle{1}}
\put(38,5.5){\circle*{1}}
\put(31,10.5){\circle*{1}}
\put(33,14){\circle*{1}}
\put(35,10.5){\circle*{1}}
\put(45,5){\makebox(0,0){$+$}}
\put(48,0){\line(1,0){18}}
\put(51,5){\line(1,0){12}}
\put(54,10){\line(1,0){6}}
\put(48,0){\line(3,5){9}}
\put(54,0){\line(3,5){6}}
\put(60,0){\line(3,5){3}}
\put(54,0){\line(-3,5){3}}
\put(60,0){\line(-3,5){6}}
\put(66,0){\line(-3,5){9}}
\put(49,0.5){\circle{1}}
\put(51,4){\circle*{1}}
\put(53,0.5){\circle{1}}
\put(55,0.5){\circle*{1}}
\put(57,4){\circle*{1}}
\put(59,0.5){\circle*{1}}
\put(61,0.5){\circle{1}}
\put(63,4){\circle*{1}}
\put(65,0.5){\circle{1}}
\put(52,5.5){\circle{1}}
\put(54,9){\circle*{1}}
\put(56,5.5){\circle{1}}
\put(58,5.5){\circle{1}}
\put(60,9){\circle*{1}}
\put(62,5.5){\circle{1}}
\put(55,10.5){\circle{1}}
\put(57,14){\circle*{1}}
\put(59,10.5){\circle{1}}
\put(69,5){\makebox(0,0){$+$}}
\put(72,0){\line(1,0){18}}
\put(75,5){\line(1,0){12}}
\put(78,10){\line(1,0){6}}
\put(72,0){\line(3,5){9}}
\put(78,0){\line(3,5){6}}
\put(84,0){\line(3,5){3}}
\put(78,0){\line(-3,5){3}}
\put(84,0){\line(-3,5){6}}
\put(90,0){\line(-3,5){9}}
\put(73,0.5){\circle{1}}
\put(75,4){\circle{1}}
\put(77,0.5){\circle*{1}}
\put(79,0.5){\circle{1}}
\put(81,4){\circle{1}}
\put(83,0.5){\circle*{1}}
\put(85,0.5){\circle{1}}
\put(87,4){\circle*{1}}
\put(89,0.5){\circle{1}}
\put(76,5.5){\circle*{1}}
\put(78,9){\circle{1}}
\put(80,5.5){\circle{1}}
\put(82,5.5){\circle*{1}}
\put(84,9){\circle{1}}
\put(86,5.5){\circle{1}}
\put(79,10.5){\circle*{1}}
\put(81,14){\circle*{1}}
\put(83,10.5){\circle*{1}}
\put(91,5){\makebox(0,0){$\times 2$}}
\put(96,5){\makebox(0,0){$+$}}
\put(99,0){\line(1,0){18}}
\put(102,5){\line(1,0){12}}
\put(105,10){\line(1,0){6}}
\put(99,0){\line(3,5){9}}
\put(105,0){\line(3,5){6}}
\put(111,0){\line(3,5){3}}
\put(105,0){\line(-3,5){3}}
\put(111,0){\line(-3,5){6}}
\put(117,0){\line(-3,5){9}}
\put(100,0.5){\circle{1}}
\put(102,4){\circle{1}}
\put(104,0.5){\circle*{1}}
\put(106,0.5){\circle{1}}
\put(108,4){\circle{1}}
\put(110,0.5){\circle*{1}}
\put(112,0.5){\circle{1}}
\put(114,4){\circle*{1}}
\put(116,0.5){\circle{1}}
\put(103,5.5){\circle*{1}}
\put(105,9){\circle*{1}}
\put(107,5.5){\circle*{1}}
\put(109,5.5){\circle{1}}
\put(111,9){\circle*{1}}
\put(113,5.5){\circle{1}}
\put(106,10.5){\circle{1}}
\put(108,14){\circle*{1}}
\put(110,10.5){\circle{1}}
\put(118,5){\makebox(0,0){$\times 2$}}
\end{picture}

\caption{\footnotesize{Illustration for the expression of $g_{2,3}(n+1)$ with positive $n$. The multiplication of two on the right-hand-side corresponds to the reflection symmetry with respect to the central vertical axis.}} 
\label{g23fig}
\end{figure}
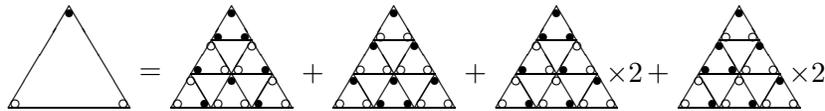

{\sl Proof} \quad 
The Sierpinski gasket $SG_{2,3}(n+1)$ is composed of six $SG_{2,3}(n)$ with six pairs of vertices identified and a set of three vertices identified. For each pair of identified vertices, either one of them is originally occupied by a dimer while the other one is vacant. For the set of three vertices, one of them is originally occupied by a dimer while the other two are vacant. The number $g_{2,3}(n+1)$ for positive $n$ consists of six configurations where five of the $SG_{2,3}(n)$ are in the $g_{2,3}(n)$ status and one in the $t_{2,3}(n)$ status as illustrated in Fig. \ref{g23fig}, such that Eq. (\ref{g23eq}) is verified. 

\bigskip

\bigskip

\begin{figure}[htbp]
\unitlength 0.9mm 
\begin{picture}(118,18)
\put(0,0){\line(1,0){18}}
\put(0,0){\line(3,5){9}}
\put(18,0){\line(-3,5){9}}
\multiput(1,0.5)(16,0){2}{\circle*{1}}
\put(9,14){\circle*{1}}
\put(21,5){\makebox(0,0){$=$}}
\put(24,0){\line(1,0){18}}
\put(27,5){\line(1,0){12}}
\put(30,10){\line(1,0){6}}
\put(24,0){\line(3,5){9}}
\put(30,0){\line(3,5){6}}
\put(36,0){\line(3,5){3}}
\put(30,0){\line(-3,5){3}}
\put(36,0){\line(-3,5){6}}
\put(42,0){\line(-3,5){9}}
\put(25,0.5){\circle*{1}}
\put(27,4){\circle*{1}}
\put(29,0.5){\circle*{1}}
\put(31,0.5){\circle{1}}
\put(33,4){\circle*{1}}
\put(35,0.5){\circle{1}}
\put(37,0.5){\circle*{1}}
\put(39,4){\circle*{1}}
\put(41,0.5){\circle*{1}}
\put(28,5.5){\circle{1}}
\put(30,9){\circle*{1}}
\put(32,5.5){\circle{1}}
\put(34,5.5){\circle{1}}
\put(36,9){\circle*{1}}
\put(38,5.5){\circle{1}}
\put(31,10.5){\circle{1}}
\put(33,14){\circle*{1}}
\put(35,10.5){\circle{1}}
\put(43,5){\makebox(0,0){$\times 3$}}
\put(48,5){\makebox(0,0){$+$}}
\put(51,0){\line(1,0){18}}
\put(54,5){\line(1,0){12}}
\put(57,10){\line(1,0){6}}
\put(51,0){\line(3,5){9}}
\put(57,0){\line(3,5){6}}
\put(63,0){\line(3,5){3}}
\put(57,0){\line(-3,5){3}}
\put(63,0){\line(-3,5){6}}
\put(69,0){\line(-3,5){9}}
\put(52,0.5){\circle*{1}}
\put(54,4){\circle{1}}
\put(56,0.5){\circle{1}}
\put(58,0.5){\circle*{1}}
\put(60,4){\circle*{1}}
\put(62,0.5){\circle*{1}}
\put(64,0.5){\circle{1}}
\put(66,4){\circle{1}}
\put(68,0.5){\circle*{1}}
\put(55,5.5){\circle*{1}}
\put(57,9){\circle{1}}
\put(59,5.5){\circle{1}}
\put(61,5.5){\circle{1}}
\put(63,9){\circle{1}}
\put(65,5.5){\circle*{1}}
\put(58,10.5){\circle*{1}}
\put(60,14){\circle*{1}}
\put(62,10.5){\circle*{1}}
\put(70,5){\makebox(0,0){$\times 3$}}
\end{picture}

\caption{\footnotesize{Illustration for the expression of $t_{2,3}(n+1)$ with positive $n$. The multiplication of three on the right-hand-side corresponds to the three possible orientations of $SG_{2,3}(n+1)$}} 
\label{t23fig}
\end{figure}
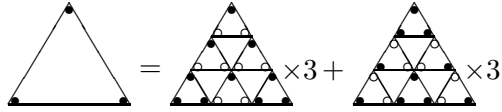

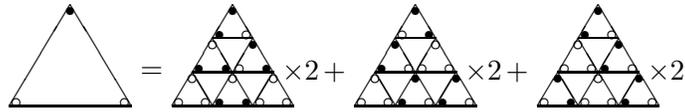
\begin{figure}[htbp]
\unitlength 0.9mm 
\begin{picture}(97,18)
\put(0,0){\line(1,0){18}}
\put(0,0){\line(3,5){9}}
\put(18,0){\line(-3,5){9}}
\multiput(1,0.5)(16,0){2}{\circle{1}}
\put(9,14){\circle*{1}}
\put(21,5){\makebox(0,0){$=$}}
\put(24,0){\line(1,0){18}}
\put(27,5){\line(1,0){12}}
\put(30,10){\line(1,0){6}}
\put(24,0){\line(3,5){9}}
\put(30,0){\line(3,5){6}}
\put(36,0){\line(3,5){3}}
\put(30,0){\line(-3,5){3}}
\put(36,0){\line(-3,5){6}}
\put(42,0){\line(-3,5){9}}
\put(25,0.5){\circle{1}}
\put(27,4){\circle{1}}
\put(29,0.5){\circle{1}}
\put(31,0.5){\circle*{1}}
\put(33,4){\circle{1}}
\put(35,0.5){\circle*{1}}
\put(37,0.5){\circle{1}}
\put(39,4){\circle{1}}
\put(41,0.5){\circle{1}}
\put(28,5.5){\circle*{1}}
\put(30,9){\circle{1}}
\put(32,5.5){\circle*{1}}
\put(34,5.5){\circle{1}}
\put(36,9){\circle*{1}}
\put(38,5.5){\circle*{1}}
\put(31,10.5){\circle*{1}}
\put(33,14){\circle*{1}}
\put(35,10.5){\circle{1}}
\put(43,5){\makebox(0,0){$\times 2$}}
\put(48,5){\makebox(0,0){$+$}}
\put(51,0){\line(1,0){18}}
\put(54,5){\line(1,0){12}}
\put(57,10){\line(1,0){6}}
\put(51,0){\line(3,5){9}}
\put(57,0){\line(3,5){6}}
\put(63,0){\line(3,5){3}}
\put(57,0){\line(-3,5){3}}
\put(63,0){\line(-3,5){6}}
\put(69,0){\line(-3,5){9}}
\put(52,0.5){\circle{1}}
\put(54,4){\circle{1}}
\put(56,0.5){\circle{1}}
\put(58,0.5){\circle*{1}}
\put(60,4){\circle*{1}}
\put(62,0.5){\circle{1}}
\put(64,0.5){\circle*{1}}
\put(66,4){\circle*{1}}
\put(68,0.5){\circle{1}}
\put(55,5.5){\circle*{1}}
\put(57,9){\circle*{1}}
\put(59,5.5){\circle{1}}
\put(61,5.5){\circle{1}}
\put(63,9){\circle{1}}
\put(65,5.5){\circle{1}}
\put(58,10.5){\circle{1}}
\put(60,14){\circle*{1}}
\put(62,10.5){\circle*{1}}
\put(70,5){\makebox(0,0){$\times 2$}}
\put(75,5){\makebox(0,0){$+$}}
\put(78,0){\line(1,0){18}}
\put(81,5){\line(1,0){12}}
\put(84,10){\line(1,0){6}}
\put(78,0){\line(3,5){9}}
\put(84,0){\line(3,5){6}}
\put(90,0){\line(3,5){3}}
\put(84,0){\line(-3,5){3}}
\put(90,0){\line(-3,5){6}}
\put(96,0){\line(-3,5){9}}
\put(79,0.5){\circle{1}}
\put(81,4){\circle*{1}}
\put(83,0.5){\circle*{1}}
\put(85,0.5){\circle{1}}
\put(87,4){\circle{1}}
\put(89,0.5){\circle{1}}
\put(91,0.5){\circle*{1}}
\put(93,4){\circle*{1}}
\put(95,0.5){\circle{1}}
\put(82,5.5){\circle{1}}
\put(84,9){\circle{1}}
\put(86,5.5){\circle{1}}
\put(88,5.5){\circle*{1}}
\put(90,9){\circle*{1}}
\put(92,5.5){\circle{1}}
\put(85,10.5){\circle*{1}}
\put(87,14){\circle*{1}}
\put(89,10.5){\circle{1}}
\put(97,5){\makebox(0,0){$\times 2$}}
\end{picture}

\caption{\footnotesize{Illustration for the expression of $g_{2,3}(1)$. The multiplication of two on the right-hand-side corresponds to the reflection symmetry with respect to the central vertical axis.}} 
\label{g231fig}
\end{figure}

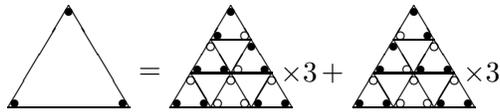
\begin{figure}[htbp]
\unitlength 0.9mm 
\begin{picture}(70,18)
\put(0,0){\line(1,0){18}}
\put(0,0){\line(3,5){9}}
\put(18,0){\line(-3,5){9}}
\multiput(1,0.5)(16,0){2}{\circle*{1}}
\put(9,14){\circle*{1}}
\put(21,5){\makebox(0,0){$=$}}
\put(24,0){\line(1,0){18}}
\put(27,5){\line(1,0){12}}
\put(30,10){\line(1,0){6}}
\put(24,0){\line(3,5){9}}
\put(30,0){\line(3,5){6}}
\put(36,0){\line(3,5){3}}
\put(30,0){\line(-3,5){3}}
\put(36,0){\line(-3,5){6}}
\put(42,0){\line(-3,5){9}}
\put(25,0.5){\circle*{1}}
\put(27,4){\circle{1}}
\put(29,0.5){\circle*{1}}
\put(31,0.5){\circle{1}}
\put(33,4){\circle{1}}
\put(35,0.5){\circle{1}}
\put(37,0.5){\circle*{1}}
\put(39,4){\circle{1}}
\put(41,0.5){\circle*{1}}
\put(28,5.5){\circle*{1}}
\put(30,9){\circle{1}}
\put(32,5.5){\circle*{1}}
\put(34,5.5){\circle{1}}
\put(36,9){\circle*{1}}
\put(38,5.5){\circle*{1}}
\put(31,10.5){\circle*{1}}
\put(33,14){\circle*{1}}
\put(35,10.5){\circle{1}}
\put(43,5){\makebox(0,0){$\times 3$}}
\put(48,5){\makebox(0,0){$+$}}
\put(51,0){\line(1,0){18}}
\put(54,5){\line(1,0){12}}
\put(57,10){\line(1,0){6}}
\put(51,0){\line(3,5){9}}
\put(57,0){\line(3,5){6}}
\put(63,0){\line(3,5){3}}
\put(57,0){\line(-3,5){3}}
\put(63,0){\line(-3,5){6}}
\put(69,0){\line(-3,5){9}}
\put(52,0.5){\circle*{1}}
\put(54,4){\circle{1}}
\put(56,0.5){\circle*{1}}
\put(58,0.5){\circle{1}}
\put(60,4){\circle{1}}
\put(62,0.5){\circle{1}}
\put(64,0.5){\circle*{1}}
\put(66,4){\circle{1}}
\put(68,0.5){\circle*{1}}
\put(55,5.5){\circle*{1}}
\put(57,9){\circle*{1}}
\put(59,5.5){\circle{1}}
\put(61,5.5){\circle*{1}}
\put(63,9){\circle{1}}
\put(65,5.5){\circle*{1}}
\put(58,10.5){\circle{1}}
\put(60,14){\circle*{1}}
\put(62,10.5){\circle*{1}}
\put(70,5){\makebox(0,0){$\times 3$}}
\end{picture}

\caption{\footnotesize{Illustration for the expression of $t_{2,3}(1)$. The multiplication of three on the right-hand-side corresponds to the three possible orientations of $SG_{2,3}(n+1)$}} 
\label{t231fig}
\end{figure}

Similarly, $t_{2,3}(n+1)$ with positive $n$ for $SG_{2,3}(n+1)$ can be obtained with appropriate configurations of its six constituting $SG_{2,3}(n)$ as illustrated in Fig. \ref{t23fig} to verify Eq. (\ref{t23eq}). Finally, $g_{2,3}(1)$ and $t_{2,3}(1)$ in Eqs. (\ref{g231eq}) and (\ref{t231eq}) are verified by Figs. \ref{g231fig} and \ref{t231fig}, respectively. \ $\Box$

\bigskip

\bigskip

\bigskip

\bigskip

It is elementary to solve $g_{2,3}(n)$ and $t_{2,3}(n)$ for positive $n$ in order to obtain the entropy for $SG_{2,3}$.

\bigskip

\begin{theo} \label{theosg23} For the generalized two-dimensional Sierpinski gasket $SG_{2,3}(n)$ at stage $n>0$,
\beq
g_{2,3}(n) = t_{2,3}(n) = 6^{\gamma_{2,3} (n)} \ ,
\label{gt23}
\eeq
where the exponent is
\beq
\gamma_{2,3} (n) = \frac15 (6^n-1) \ .
\label{alpha23}
\eeq
Define the number of dimer coverings $N(SG_{2,3}(n))$ in Eq. (\ref{sdef}) equal to $t_{2,3}(n)$. With $v(SG_{2,3}(n))=(7(6)^n+8)/5$, the entropy is given by 
\beq
S_{SG_{2,3}} = \frac17 \ln 6 \simeq 0.25596563846...
\label{Ssg23}
\eeq
\end{theo}

\bigskip

\subsection{$SG_{2,4}(n)$}

For the generalized two-dimensional Sierpinski gasket $SG_{2,b}(n)$ with $b=4$, we have
\beq
v(SG_{2,4}(n)) = \frac{4(10)^n+5}{3} = 3 + \frac43 (10^n-1) 
= 3 + \frac43 \sum_{j=1}^n {n \choose j} 9^j
\eeq
by Eq. (\ref{bv}), such that the number of vertices is always odd for any $n$. Therefore, $g_{2,4}(n)$ and $t_{2,4}(n)$ are zero for all $n$, while the initial values remain $f_{2,4}(0)=1$, $g_{2,4}(0)=0$, $h_{2,4}(0)=1$ and $t_{2,4}(0)=0$. $f_{2,4}(n)$ and $h_{2,4}(n)$ satisfy recursion relations.

\bigskip

\begin{lemma} \label{lemmasg24r} For any non-negative integer $n$,
\beq
f_{2,4}(n+1) = 28 f_{2,4}^4(n) h_{2,4}^6(n) \ , 
\label{f24eq}
\eeq
\beq
h_{2,4}(n+1) = 28 f_{2,4}^3(n) h_{2,4}^7(n) \ .
\label{h24eq}
\eeq
\end{lemma}

\begin{figure}[htbp]
\unitlength 0.9mm 
\begin{picture}(121,24)
\put(0,0){\line(1,0){24}}
\put(0,0){\line(3,5){12}}
\put(24,0){\line(-3,5){12}}
\multiput(1,0.5)(22,0){2}{\circle{1}}
\put(12,19){\circle{1}}
\put(27,5){\makebox(0,0){$=$}}
\put(30,0){\line(1,0){24}}
\put(33,5){\line(1,0){18}}
\put(36,10){\line(1,0){12}}
\put(39,15){\line(1,0){6}}
\put(30,0){\line(3,5){12}}
\put(36,0){\line(3,5){9}}
\put(42,0){\line(3,5){6}}
\put(48,0){\line(3,5){3}}
\put(36,0){\line(-3,5){3}}
\put(42,0){\line(-3,5){6}}
\put(48,0){\line(-3,5){9}}
\put(54,0){\line(-3,5){12}}
\put(31,0.5){\circle{1}}
\put(33,4){\circle{1}}
\put(35,0.5){\circle{1}}
\put(37,0.5){\circle*{1}}
\put(39,4){\circle{1}}
\put(41,0.5){\circle*{1}}
\put(43,0.5){\circle{1}}
\put(45,4){\circle*{1}}
\put(47,0.5){\circle*{1}}
\put(49,0.5){\circle{1}}
\put(51,4){\circle{1}}
\put(53,0.5){\circle{1}}
\put(34,5.5){\circle*{1}}
\put(36,9){\circle{1}}
\put(38,5.5){\circle*{1}}
\put(40,5.5){\circle{1}}
\put(42,9){\circle{1}}
\put(44,5.5){\circle{1}}
\put(46,5.5){\circle{1}}
\put(48,9){\circle*{1}}
\put(50,5.5){\circle*{1}}
\put(37,10.5){\circle*{1}}
\put(39,14){\circle*{1}}
\put(41,10.5){\circle{1}}
\put(43,10.5){\circle*{1}}
\put(45,14){\circle*{1}}
\put(47,10.5){\circle{1}}
\put(40,15.5){\circle{1}}
\put(42,19){\circle{1}}
\put(44,15.5){\circle{1}}
\put(60,5){\makebox(0,0){$+$}}
\put(63,0){\line(1,0){24}}
\put(66,5){\line(1,0){18}}
\put(69,10){\line(1,0){12}}
\put(72,15){\line(1,0){6}}
\put(63,0){\line(3,5){12}}
\put(69,0){\line(3,5){9}}
\put(75,0){\line(3,5){6}}
\put(81,0){\line(3,5){3}}
\put(69,0){\line(-3,5){3}}
\put(75,0){\line(-3,5){6}}
\put(81,0){\line(-3,5){9}}
\put(87,0){\line(-3,5){12}}
\put(64,0.5){\circle{1}}
\put(66,4){\circle{1}}
\put(68,0.5){\circle{1}}
\put(70,0.5){\circle*{1}}
\put(72,4){\circle*{1}}
\put(74,0.5){\circle{1}}
\put(76,0.5){\circle*{1}}
\put(78,4){\circle{1}}
\put(80,0.5){\circle*{1}}
\put(82,0.5){\circle{1}}
\put(84,4){\circle{1}}
\put(86,0.5){\circle{1}}
\put(67,5.5){\circle*{1}}
\put(69,9){\circle*{1}}
\put(71,5.5){\circle{1}}
\put(73,5.5){\circle{1}}
\put(75,9){\circle{1}}
\put(77,5.5){\circle{1}}
\put(79,5.5){\circle*{1}}
\put(81,9){\circle{1}}
\put(83,5.5){\circle*{1}}
\put(70,10.5){\circle{1}}
\put(72,14){\circle*{1}}
\put(74,10.5){\circle*{1}}
\put(76,10.5){\circle{1}}
\put(78,14){\circle*{1}}
\put(80,10.5){\circle*{1}}
\put(73,15.5){\circle{1}}
\put(75,19){\circle{1}}
\put(77,15.5){\circle{1}}
\put(93,5){\makebox(0,0){$+$}}
\put(96,0){\line(1,0){24}}
\put(99,5){\line(1,0){18}}
\put(102,10){\line(1,0){12}}
\put(105,15){\line(1,0){6}}
\put(96,0){\line(3,5){12}}
\put(102,0){\line(3,5){9}}
\put(108,0){\line(3,5){6}}
\put(114,0){\line(3,5){3}}
\put(102,0){\line(-3,5){3}}
\put(108,0){\line(-3,5){6}}
\put(114,0){\line(-3,5){9}}
\put(120,0){\line(-3,5){12}}
\put(97,0.5){\circle{1}}
\put(99,4){\circle*{1}}
\put(101,0.5){\circle*{1}}
\put(103,0.5){\circle{1}}
\put(105,4){\circle{1}}
\put(107,0.5){\circle{1}}
\put(109,0.5){\circle*{1}}
\put(111,4){\circle*{1}}
\put(113,0.5){\circle{1}}
\put(115,0.5){\circle*{1}}
\put(117,4){\circle*{1}}
\put(119,0.5){\circle{1}}
\put(100,5.5){\circle{1}}
\put(102,9){\circle*{1}}
\put(104,5.5){\circle*{1}}
\put(106,5.5){\circle{1}}
\put(108,9){\circle{1}}
\put(110,5.5){\circle{1}}
\put(112,5.5){\circle{1}}
\put(114,9){\circle{1}}
\put(116,5.5){\circle{1}}
\put(103,10.5){\circle{1}}
\put(105,14){\circle{1}}
\put(107,10.5){\circle{1}}
\put(109,10.5){\circle*{1}}
\put(111,14){\circle{1}}
\put(113,10.5){\circle*{1}}
\put(106,15.5){\circle*{1}}
\put(108,19){\circle{1}}
\put(110,15.5){\circle*{1}}
\end{picture}

\begin{picture}(121,24)
\put(27,5){\makebox(0,0){$+$}}
\put(30,0){\line(1,0){24}}
\put(33,5){\line(1,0){18}}
\put(36,10){\line(1,0){12}}
\put(39,15){\line(1,0){6}}
\put(30,0){\line(3,5){12}}
\put(36,0){\line(3,5){9}}
\put(42,0){\line(3,5){6}}
\put(48,0){\line(3,5){3}}
\put(36,0){\line(-3,5){3}}
\put(42,0){\line(-3,5){6}}
\put(48,0){\line(-3,5){9}}
\put(54,0){\line(-3,5){12}}
\put(31,0.5){\circle{1}}
\put(33,4){\circle*{1}}
\put(35,0.5){\circle*{1}}
\put(37,0.5){\circle{1}}
\put(39,4){\circle*{1}}
\put(41,0.5){\circle*{1}}
\put(43,0.5){\circle{1}}
\put(45,4){\circle{1}}
\put(47,0.5){\circle{1}}
\put(49,0.5){\circle*{1}}
\put(51,4){\circle*{1}}
\put(53,0.5){\circle{1}}
\put(34,5.5){\circle{1}}
\put(36,9){\circle{1}}
\put(38,5.5){\circle{1}}
\put(40,5.5){\circle{1}}
\put(42,9){\circle{1}}
\put(44,5.5){\circle{1}}
\put(46,5.5){\circle*{1}}
\put(48,9){\circle*{1}}
\put(50,5.5){\circle{1}}
\put(37,10.5){\circle*{1}}
\put(39,14){\circle{1}}
\put(41,10.5){\circle*{1}}
\put(43,10.5){\circle{1}}
\put(45,14){\circle{1}}
\put(47,10.5){\circle{1}}
\put(40,15.5){\circle*{1}}
\put(42,19){\circle{1}}
\put(44,15.5){\circle*{1}}
\put(60,5){\makebox(0,0){$+$}}
\put(63,0){\line(1,0){24}}
\put(66,5){\line(1,0){18}}
\put(69,10){\line(1,0){12}}
\put(72,15){\line(1,0){6}}
\put(63,0){\line(3,5){12}}
\put(69,0){\line(3,5){9}}
\put(75,0){\line(3,5){6}}
\put(81,0){\line(3,5){3}}
\put(69,0){\line(-3,5){3}}
\put(75,0){\line(-3,5){6}}
\put(81,0){\line(-3,5){9}}
\put(87,0){\line(-3,5){12}}
\put(64,0.5){\circle{1}}
\put(66,4){\circle{1}}
\put(68,0.5){\circle{1}}
\put(70,0.5){\circle*{1}}
\put(72,4){\circle{1}}
\put(74,0.5){\circle*{1}}
\put(76,0.5){\circle{1}}
\put(78,4){\circle*{1}}
\put(80,0.5){\circle*{1}}
\put(82,0.5){\circle{1}}
\put(84,4){\circle{1}}
\put(86,0.5){\circle{1}}
\put(67,5.5){\circle*{1}}
\put(69,9){\circle*{1}}
\put(71,5.5){\circle{1}}
\put(73,5.5){\circle*{1}}
\put(75,9){\circle*{1}}
\put(77,5.5){\circle{1}}
\put(79,5.5){\circle{1}}
\put(81,9){\circle*{1}}
\put(83,5.5){\circle*{1}}
\put(70,10.5){\circle{1}}
\put(72,14){\circle{1}}
\put(74,10.5){\circle{1}}
\put(76,10.5){\circle{1}}
\put(78,14){\circle{1}}
\put(80,10.5){\circle{1}}
\put(73,15.5){\circle*{1}}
\put(75,19){\circle{1}}
\put(77,15.5){\circle*{1}}
\put(88,5){\makebox(0,0){$\times 3$}}
\put(93,5){\makebox(0,0){$+$}}
\put(96,0){\line(1,0){24}}
\put(99,5){\line(1,0){18}}
\put(102,10){\line(1,0){12}}
\put(105,15){\line(1,0){6}}
\put(96,0){\line(3,5){12}}
\put(102,0){\line(3,5){9}}
\put(108,0){\line(3,5){6}}
\put(114,0){\line(3,5){3}}
\put(102,0){\line(-3,5){3}}
\put(108,0){\line(-3,5){6}}
\put(114,0){\line(-3,5){9}}
\put(120,0){\line(-3,5){12}}
\put(97,0.5){\circle{1}}
\put(99,4){\circle{1}}
\put(101,0.5){\circle{1}}
\put(103,0.5){\circle*{1}}
\put(105,4){\circle*{1}}
\put(107,0.5){\circle{1}}
\put(109,0.5){\circle*{1}}
\put(111,4){\circle{1}}
\put(113,0.5){\circle*{1}}
\put(115,0.5){\circle{1}}
\put(117,4){\circle{1}}
\put(119,0.5){\circle{1}}
\put(100,5.5){\circle*{1}}
\put(102,9){\circle*{1}}
\put(104,5.5){\circle{1}}
\put(106,5.5){\circle{1}}
\put(108,9){\circle*{1}}
\put(110,5.5){\circle*{1}}
\put(112,5.5){\circle{1}}
\put(114,9){\circle*{1}}
\put(116,5.5){\circle*{1}}
\put(103,10.5){\circle{1}}
\put(105,14){\circle{1}}
\put(107,10.5){\circle{1}}
\put(109,10.5){\circle{1}}
\put(111,14){\circle{1}}
\put(113,10.5){\circle{1}}
\put(106,15.5){\circle*{1}}
\put(108,19){\circle{1}}
\put(110,15.5){\circle*{1}}
\put(121,5){\makebox(0,0){$\times 3$}}
\end{picture}

\begin{picture}(121,24)
\put(27,5){\makebox(0,0){$+$}}
\put(30,0){\line(1,0){24}}
\put(33,5){\line(1,0){18}}
\put(36,10){\line(1,0){12}}
\put(39,15){\line(1,0){6}}
\put(30,0){\line(3,5){12}}
\put(36,0){\line(3,5){9}}
\put(42,0){\line(3,5){6}}
\put(48,0){\line(3,5){3}}
\put(36,0){\line(-3,5){3}}
\put(42,0){\line(-3,5){6}}
\put(48,0){\line(-3,5){9}}
\put(54,0){\line(-3,5){12}}
\put(31,0.5){\circle{1}}
\put(33,4){\circle{1}}
\put(35,0.5){\circle{1}}
\put(37,0.5){\circle*{1}}
\put(39,4){\circle*{1}}
\put(41,0.5){\circle{1}}
\put(43,0.5){\circle*{1}}
\put(45,4){\circle{1}}
\put(47,0.5){\circle*{1}}
\put(49,0.5){\circle{1}}
\put(51,4){\circle{1}}
\put(53,0.5){\circle{1}}
\put(34,5.5){\circle*{1}}
\put(36,9){\circle*{1}}
\put(38,5.5){\circle{1}}
\put(40,5.5){\circle{1}}
\put(42,9){\circle{1}}
\put(44,5.5){\circle{1}}
\put(46,5.5){\circle*{1}}
\put(48,9){\circle{1}}
\put(50,5.5){\circle*{1}}
\put(37,10.5){\circle{1}}
\put(39,14){\circle{1}}
\put(41,10.5){\circle{1}}
\put(43,10.5){\circle*{1}}
\put(45,14){\circle{1}}
\put(47,10.5){\circle*{1}}
\put(40,15.5){\circle*{1}}
\put(42,19){\circle{1}}
\put(44,15.5){\circle*{1}}
\put(55,5){\makebox(0,0){$\times 3$}}
\put(60,5){\makebox(0,0){$+$}}
\put(63,0){\line(1,0){24}}
\put(66,5){\line(1,0){18}}
\put(69,10){\line(1,0){12}}
\put(72,15){\line(1,0){6}}
\put(63,0){\line(3,5){12}}
\put(69,0){\line(3,5){9}}
\put(75,0){\line(3,5){6}}
\put(81,0){\line(3,5){3}}
\put(69,0){\line(-3,5){3}}
\put(75,0){\line(-3,5){6}}
\put(81,0){\line(-3,5){9}}
\put(87,0){\line(-3,5){12}}
\put(64,0.5){\circle{1}}
\put(66,4){\circle{1}}
\put(68,0.5){\circle{1}}
\put(70,0.5){\circle*{1}}
\put(72,4){\circle{1}}
\put(74,0.5){\circle*{1}}
\put(76,0.5){\circle{1}}
\put(78,4){\circle*{1}}
\put(80,0.5){\circle*{1}}
\put(82,0.5){\circle{1}}
\put(84,4){\circle{1}}
\put(86,0.5){\circle{1}}
\put(67,5.5){\circle*{1}}
\put(69,9){\circle{1}}
\put(71,5.5){\circle*{1}}
\put(73,5.5){\circle{1}}
\put(75,9){\circle{1}}
\put(77,5.5){\circle{1}}
\put(79,5.5){\circle{1}}
\put(81,9){\circle*{1}}
\put(83,5.5){\circle*{1}}
\put(70,10.5){\circle*{1}}
\put(72,14){\circle{1}}
\put(74,10.5){\circle*{1}}
\put(76,10.5){\circle{1}}
\put(78,14){\circle{1}}
\put(80,10.5){\circle{1}}
\put(73,15.5){\circle*{1}}
\put(75,19){\circle{1}}
\put(77,15.5){\circle*{1}}
\put(88,5){\makebox(0,0){$\times 3$}}
\put(93,5){\makebox(0,0){$+$}}
\put(96,0){\line(1,0){24}}
\put(99,5){\line(1,0){18}}
\put(102,10){\line(1,0){12}}
\put(105,15){\line(1,0){6}}
\put(96,0){\line(3,5){12}}
\put(102,0){\line(3,5){9}}
\put(108,0){\line(3,5){6}}
\put(114,0){\line(3,5){3}}
\put(102,0){\line(-3,5){3}}
\put(108,0){\line(-3,5){6}}
\put(114,0){\line(-3,5){9}}
\put(120,0){\line(-3,5){12}}
\put(97,0.5){\circle{1}}
\put(99,4){\circle*{1}}
\put(101,0.5){\circle*{1}}
\put(103,0.5){\circle{1}}
\put(105,4){\circle{1}}
\put(107,0.5){\circle{1}}
\put(109,0.5){\circle*{1}}
\put(111,4){\circle{1}}
\put(113,0.5){\circle*{1}}
\put(115,0.5){\circle{1}}
\put(117,4){\circle{1}}
\put(119,0.5){\circle{1}}
\put(100,5.5){\circle{1}}
\put(102,9){\circle*{1}}
\put(104,5.5){\circle*{1}}
\put(106,5.5){\circle{1}}
\put(108,9){\circle*{1}}
\put(110,5.5){\circle*{1}}
\put(112,5.5){\circle{1}}
\put(114,9){\circle*{1}}
\put(116,5.5){\circle*{1}}
\put(103,10.5){\circle{1}}
\put(105,14){\circle{1}}
\put(107,10.5){\circle{1}}
\put(109,10.5){\circle{1}}
\put(111,14){\circle{1}}
\put(113,10.5){\circle{1}}
\put(106,15.5){\circle*{1}}
\put(108,19){\circle{1}}
\put(110,15.5){\circle*{1}}
\put(121,5){\makebox(0,0){$\times 3$}}
\end{picture}

\begin{picture}(121,24)
\put(27,5){\makebox(0,0){$+$}}
\put(30,0){\line(1,0){24}}
\put(33,5){\line(1,0){18}}
\put(36,10){\line(1,0){12}}
\put(39,15){\line(1,0){6}}
\put(30,0){\line(3,5){12}}
\put(36,0){\line(3,5){9}}
\put(42,0){\line(3,5){6}}
\put(48,0){\line(3,5){3}}
\put(36,0){\line(-3,5){3}}
\put(42,0){\line(-3,5){6}}
\put(48,0){\line(-3,5){9}}
\put(54,0){\line(-3,5){12}}
\put(31,0.5){\circle{1}}
\put(33,4){\circle{1}}
\put(35,0.5){\circle{1}}
\put(37,0.5){\circle*{1}}
\put(39,4){\circle{1}}
\put(41,0.5){\circle*{1}}
\put(43,0.5){\circle{1}}
\put(45,4){\circle{1}}
\put(47,0.5){\circle{1}}
\put(49,0.5){\circle*{1}}
\put(51,4){\circle*{1}}
\put(53,0.5){\circle{1}}
\put(34,5.5){\circle*{1}}
\put(36,9){\circle*{1}}
\put(38,5.5){\circle{1}}
\put(40,5.5){\circle*{1}}
\put(42,9){\circle*{1}}
\put(44,5.5){\circle{1}}
\put(46,5.5){\circle*{1}}
\put(48,9){\circle*{1}}
\put(50,5.5){\circle{1}}
\put(37,10.5){\circle{1}}
\put(39,14){\circle{1}}
\put(41,10.5){\circle{1}}
\put(43,10.5){\circle{1}}
\put(45,14){\circle{1}}
\put(47,10.5){\circle{1}}
\put(40,15.5){\circle*{1}}
\put(42,19){\circle{1}}
\put(44,15.5){\circle*{1}}
\put(55,5){\makebox(0,0){$\times 3$}}
\put(60,5){\makebox(0,0){$+$}}
\put(63,0){\line(1,0){24}}
\put(66,5){\line(1,0){18}}
\put(69,10){\line(1,0){12}}
\put(72,15){\line(1,0){6}}
\put(63,0){\line(3,5){12}}
\put(69,0){\line(3,5){9}}
\put(75,0){\line(3,5){6}}
\put(81,0){\line(3,5){3}}
\put(69,0){\line(-3,5){3}}
\put(75,0){\line(-3,5){6}}
\put(81,0){\line(-3,5){9}}
\put(87,0){\line(-3,5){12}}
\put(64,0.5){\circle{1}}
\put(66,4){\circle*{1}}
\put(68,0.5){\circle*{1}}
\put(70,0.5){\circle{1}}
\put(72,4){\circle{1}}
\put(74,0.5){\circle{1}}
\put(76,0.5){\circle*{1}}
\put(78,4){\circle{1}}
\put(80,0.5){\circle*{1}}
\put(82,0.5){\circle{1}}
\put(84,4){\circle{1}}
\put(86,0.5){\circle{1}}
\put(67,5.5){\circle{1}}
\put(69,9){\circle*{1}}
\put(71,5.5){\circle*{1}}
\put(73,5.5){\circle{1}}
\put(75,9){\circle{1}}
\put(77,5.5){\circle{1}}
\put(79,5.5){\circle*{1}}
\put(81,9){\circle{1}}
\put(83,5.5){\circle*{1}}
\put(70,10.5){\circle{1}}
\put(72,14){\circle{1}}
\put(74,10.5){\circle{1}}
\put(76,10.5){\circle*{1}}
\put(78,14){\circle{1}}
\put(80,10.5){\circle*{1}}
\put(73,15.5){\circle*{1}}
\put(75,19){\circle{1}}
\put(77,15.5){\circle*{1}}
\put(88,5){\makebox(0,0){$\times 3$}}
\put(93,5){\makebox(0,0){$+$}}
\put(96,0){\line(1,0){24}}
\put(99,5){\line(1,0){18}}
\put(102,10){\line(1,0){12}}
\put(105,15){\line(1,0){6}}
\put(96,0){\line(3,5){12}}
\put(102,0){\line(3,5){9}}
\put(108,0){\line(3,5){6}}
\put(114,0){\line(3,5){3}}
\put(102,0){\line(-3,5){3}}
\put(108,0){\line(-3,5){6}}
\put(114,0){\line(-3,5){9}}
\put(120,0){\line(-3,5){12}}
\put(97,0.5){\circle{1}}
\put(99,4){\circle{1}}
\put(101,0.5){\circle{1}}
\put(103,0.5){\circle*{1}}
\put(105,4){\circle{1}}
\put(107,0.5){\circle*{1}}
\put(109,0.5){\circle{1}}
\put(111,4){\circle{1}}
\put(113,0.5){\circle{1}}
\put(115,0.5){\circle*{1}}
\put(117,4){\circle*{1}}
\put(119,0.5){\circle{1}}
\put(100,5.5){\circle*{1}}
\put(102,9){\circle{1}}
\put(104,5.5){\circle*{1}}
\put(106,5.5){\circle{1}}
\put(108,9){\circle{1}}
\put(110,5.5){\circle{1}}
\put(112,5.5){\circle*{1}}
\put(114,9){\circle*{1}}
\put(116,5.5){\circle{1}}
\put(103,10.5){\circle*{1}}
\put(105,14){\circle{1}}
\put(107,10.5){\circle*{1}}
\put(109,10.5){\circle{1}}
\put(111,14){\circle{1}}
\put(113,10.5){\circle{1}}
\put(106,15.5){\circle*{1}}
\put(108,19){\circle{1}}
\put(110,15.5){\circle*{1}}
\put(121,5){\makebox(0,0){$\times 3$}}
\end{picture}

\caption{\footnotesize{Illustration for the expression of $f_{2,4}(n+1)$ with non-negative $n$. The multiplication of three on the right-hand-side corresponds to the three possible orientations of $SG_{2,4}(n+1)$.}} 
\label{f24fig}
\end{figure}
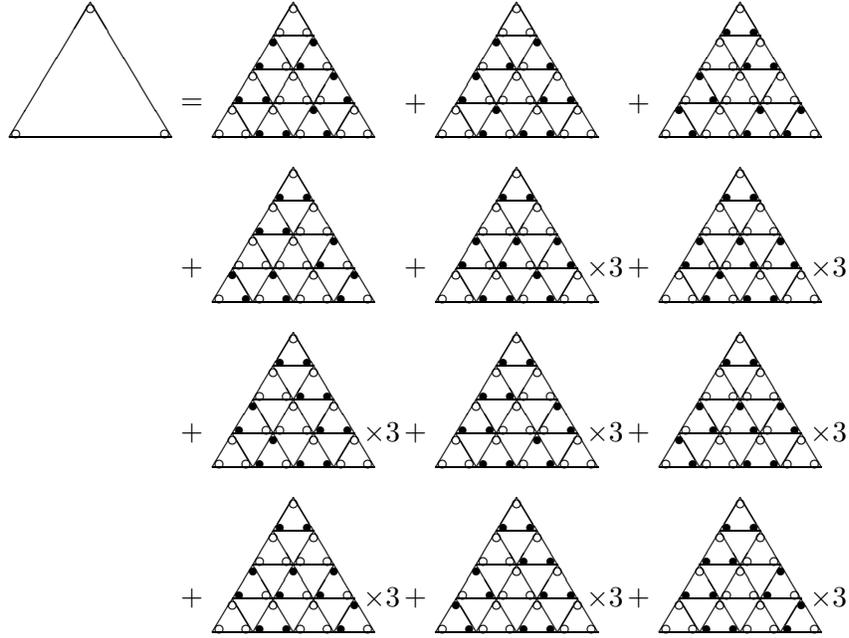

{\sl Proof} \quad 
The Sierpinski gasket $SG_{2,4}(n+1)$ is composed of ten $SG_{2,4}(n)$ with nine pairs of vertices identified and three sets of three vertices identified. The number $f_{2,4}(n+1)$ for non-negative $n$ consists of twenty eight configurations where four of the $SG_{2,4}(n)$ are in the $f_{2,4}(n)$ status and six in the $h_{2,4}(n)$ status as illustrated in Fig. \ref{f24fig}, such that Eq. (\ref{f24eq}) is verified. 

\bigskip

\bigskip

\begin{figure}[htbp]
\unitlength 0.9mm 
\begin{picture}(121,24)
\put(0,0){\line(1,0){24}}
\put(0,0){\line(3,5){12}}
\put(24,0){\line(-3,5){12}}
\multiput(1,0.5)(22,0){2}{\circle*{1}}
\put(12,19){\circle{1}}
\put(27,5){\makebox(0,0){$=$}}
\put(30,0){\line(1,0){24}}
\put(33,5){\line(1,0){18}}
\put(36,10){\line(1,0){12}}
\put(39,15){\line(1,0){6}}
\put(30,0){\line(3,5){12}}
\put(36,0){\line(3,5){9}}
\put(42,0){\line(3,5){6}}
\put(48,0){\line(3,5){3}}
\put(36,0){\line(-3,5){3}}
\put(42,0){\line(-3,5){6}}
\put(48,0){\line(-3,5){9}}
\put(54,0){\line(-3,5){12}}
\put(31,0.5){\circle*{1}}
\put(33,4){\circle*{1}}
\put(35,0.5){\circle{1}}
\put(37,0.5){\circle*{1}}
\put(39,4){\circle{1}}
\put(41,0.5){\circle*{1}}
\put(43,0.5){\circle{1}}
\put(45,4){\circle*{1}}
\put(47,0.5){\circle*{1}}
\put(49,0.5){\circle{1}}
\put(51,4){\circle*{1}}
\put(53,0.5){\circle*{1}}
\put(34,5.5){\circle{1}}
\put(36,9){\circle{1}}
\put(38,5.5){\circle{1}}
\put(40,5.5){\circle*{1}}
\put(42,9){\circle*{1}}
\put(44,5.5){\circle{1}}
\put(46,5.5){\circle{1}}
\put(48,9){\circle{1}}
\put(50,5.5){\circle{1}}
\put(37,10.5){\circle*{1}}
\put(39,14){\circle*{1}}
\put(41,10.5){\circle{1}}
\put(43,10.5){\circle{1}}
\put(45,14){\circle*{1}}
\put(47,10.5){\circle*{1}}
\put(40,15.5){\circle{1}}
\put(42,19){\circle{1}}
\put(44,15.5){\circle{1}}
\put(55,5){\makebox(0,0){$\times 2$}}
\put(60,5){\makebox(0,0){$+$}}
\put(63,0){\line(1,0){24}}
\put(66,5){\line(1,0){18}}
\put(69,10){\line(1,0){12}}
\put(72,15){\line(1,0){6}}
\put(63,0){\line(3,5){12}}
\put(69,0){\line(3,5){9}}
\put(75,0){\line(3,5){6}}
\put(81,0){\line(3,5){3}}
\put(69,0){\line(-3,5){3}}
\put(75,0){\line(-3,5){6}}
\put(81,0){\line(-3,5){9}}
\put(87,0){\line(-3,5){12}}
\put(64,0.5){\circle*{1}}
\put(66,4){\circle*{1}}
\put(68,0.5){\circle{1}}
\put(70,0.5){\circle*{1}}
\put(72,4){\circle{1}}
\put(74,0.5){\circle*{1}}
\put(76,0.5){\circle{1}}
\put(78,4){\circle{1}}
\put(80,0.5){\circle{1}}
\put(82,0.5){\circle*{1}}
\put(84,4){\circle{1}}
\put(86,0.5){\circle*{1}}
\put(67,5.5){\circle{1}}
\put(69,9){\circle{1}}
\put(71,5.5){\circle{1}}
\put(73,5.5){\circle*{1}}
\put(75,9){\circle{1}}
\put(77,5.5){\circle*{1}}
\put(79,5.5){\circle{1}}
\put(81,9){\circle*{1}}
\put(83,5.5){\circle*{1}}
\put(70,10.5){\circle*{1}}
\put(72,14){\circle{1}}
\put(74,10.5){\circle*{1}}
\put(76,10.5){\circle{1}}
\put(78,14){\circle{1}}
\put(80,10.5){\circle{1}}
\put(73,15.5){\circle*{1}}
\put(75,19){\circle{1}}
\put(77,15.5){\circle*{1}}
\put(88,5){\makebox(0,0){$\times 2$}}
\put(93,5){\makebox(0,0){$+$}}
\put(96,0){\line(1,0){24}}
\put(99,5){\line(1,0){18}}
\put(102,10){\line(1,0){12}}
\put(105,15){\line(1,0){6}}
\put(96,0){\line(3,5){12}}
\put(102,0){\line(3,5){9}}
\put(108,0){\line(3,5){6}}
\put(114,0){\line(3,5){3}}
\put(102,0){\line(-3,5){3}}
\put(108,0){\line(-3,5){6}}
\put(114,0){\line(-3,5){9}}
\put(120,0){\line(-3,5){12}}
\put(97,0.5){\circle*{1}}
\put(99,4){\circle*{1}}
\put(101,0.5){\circle{1}}
\put(103,0.5){\circle*{1}}
\put(105,4){\circle*{1}}
\put(107,0.5){\circle{1}}
\put(109,0.5){\circle*{1}}
\put(111,4){\circle{1}}
\put(113,0.5){\circle*{1}}
\put(115,0.5){\circle{1}}
\put(117,4){\circle*{1}}
\put(119,0.5){\circle*{1}}
\put(100,5.5){\circle{1}}
\put(102,9){\circle{1}}
\put(104,5.5){\circle{1}}
\put(106,5.5){\circle{1}}
\put(108,9){\circle{1}}
\put(110,5.5){\circle{1}}
\put(112,5.5){\circle*{1}}
\put(114,9){\circle*{1}}
\put(116,5.5){\circle{1}}
\put(103,10.5){\circle*{1}}
\put(105,14){\circle*{1}}
\put(107,10.5){\circle{1}}
\put(109,10.5){\circle*{1}}
\put(111,14){\circle*{1}}
\put(113,10.5){\circle{1}}
\put(106,15.5){\circle{1}}
\put(108,19){\circle{1}}
\put(110,15.5){\circle{1}}
\put(121,5){\makebox(0,0){$\times 2$}}
\end{picture}

\begin{picture}(121,24)
\put(27,5){\makebox(0,0){$+$}}
\put(30,0){\line(1,0){24}}
\put(33,5){\line(1,0){18}}
\put(36,10){\line(1,0){12}}
\put(39,15){\line(1,0){6}}
\put(30,0){\line(3,5){12}}
\put(36,0){\line(3,5){9}}
\put(42,0){\line(3,5){6}}
\put(48,0){\line(3,5){3}}
\put(36,0){\line(-3,5){3}}
\put(42,0){\line(-3,5){6}}
\put(48,0){\line(-3,5){9}}
\put(54,0){\line(-3,5){12}}
\put(31,0.5){\circle*{1}}
\put(33,4){\circle*{1}}
\put(35,0.5){\circle{1}}
\put(37,0.5){\circle*{1}}
\put(39,4){\circle*{1}}
\put(41,0.5){\circle{1}}
\put(43,0.5){\circle*{1}}
\put(45,4){\circle*{1}}
\put(47,0.5){\circle{1}}
\put(49,0.5){\circle*{1}}
\put(51,4){\circle{1}}
\put(53,0.5){\circle*{1}}
\put(34,5.5){\circle{1}}
\put(36,9){\circle{1}}
\put(38,5.5){\circle{1}}
\put(40,5.5){\circle{1}}
\put(42,9){\circle{1}}
\put(44,5.5){\circle{1}}
\put(46,5.5){\circle{1}}
\put(48,9){\circle*{1}}
\put(50,5.5){\circle*{1}}
\put(37,10.5){\circle*{1}}
\put(39,14){\circle*{1}}
\put(41,10.5){\circle{1}}
\put(43,10.5){\circle*{1}}
\put(45,14){\circle*{1}}
\put(47,10.5){\circle{1}}
\put(40,15.5){\circle{1}}
\put(42,19){\circle{1}}
\put(44,15.5){\circle{1}}
\put(55,5){\makebox(0,0){$\times 2$}}
\put(60,5){\makebox(0,0){$+$}}
\put(63,0){\line(1,0){24}}
\put(66,5){\line(1,0){18}}
\put(69,10){\line(1,0){12}}
\put(72,15){\line(1,0){6}}
\put(63,0){\line(3,5){12}}
\put(69,0){\line(3,5){9}}
\put(75,0){\line(3,5){6}}
\put(81,0){\line(3,5){3}}
\put(69,0){\line(-3,5){3}}
\put(75,0){\line(-3,5){6}}
\put(81,0){\line(-3,5){9}}
\put(87,0){\line(-3,5){12}}
\put(64,0.5){\circle*{1}}
\put(66,4){\circle{1}}
\put(68,0.5){\circle*{1}}
\put(70,0.5){\circle{1}}
\put(72,4){\circle{1}}
\put(74,0.5){\circle{1}}
\put(76,0.5){\circle*{1}}
\put(78,4){\circle{1}}
\put(80,0.5){\circle*{1}}
\put(82,0.5){\circle{1}}
\put(84,4){\circle*{1}}
\put(86,0.5){\circle*{1}}
\put(67,5.5){\circle*{1}}
\put(69,9){\circle{1}}
\put(71,5.5){\circle*{1}}
\put(73,5.5){\circle{1}}
\put(75,9){\circle{1}}
\put(77,5.5){\circle{1}}
\put(79,5.5){\circle*{1}}
\put(81,9){\circle*{1}}
\put(83,5.5){\circle{1}}
\put(70,10.5){\circle*{1}}
\put(72,14){\circle*{1}}
\put(74,10.5){\circle{1}}
\put(76,10.5){\circle*{1}}
\put(78,14){\circle*{1}}
\put(80,10.5){\circle{1}}
\put(73,15.5){\circle{1}}
\put(75,19){\circle{1}}
\put(77,15.5){\circle{1}}
\put(88,5){\makebox(0,0){$\times 2$}}
\put(93,5){\makebox(0,0){$+$}}
\put(96,0){\line(1,0){24}}
\put(99,5){\line(1,0){18}}
\put(102,10){\line(1,0){12}}
\put(105,15){\line(1,0){6}}
\put(96,0){\line(3,5){12}}
\put(102,0){\line(3,5){9}}
\put(108,0){\line(3,5){6}}
\put(114,0){\line(3,5){3}}
\put(102,0){\line(-3,5){3}}
\put(108,0){\line(-3,5){6}}
\put(114,0){\line(-3,5){9}}
\put(120,0){\line(-3,5){12}}
\put(97,0.5){\circle*{1}}
\put(99,4){\circle{1}}
\put(101,0.5){\circle*{1}}
\put(103,0.5){\circle{1}}
\put(105,4){\circle{1}}
\put(107,0.5){\circle{1}}
\put(109,0.5){\circle*{1}}
\put(111,4){\circle*{1}}
\put(113,0.5){\circle{1}}
\put(115,0.5){\circle*{1}}
\put(117,4){\circle{1}}
\put(119,0.5){\circle*{1}}
\put(100,5.5){\circle*{1}}
\put(102,9){\circle{1}}
\put(104,5.5){\circle*{1}}
\put(106,5.5){\circle{1}}
\put(108,9){\circle{1}}
\put(110,5.5){\circle{1}}
\put(112,5.5){\circle{1}}
\put(114,9){\circle*{1}}
\put(116,5.5){\circle*{1}}
\put(103,10.5){\circle*{1}}
\put(105,14){\circle*{1}}
\put(107,10.5){\circle{1}}
\put(109,10.5){\circle*{1}}
\put(111,14){\circle*{1}}
\put(113,10.5){\circle{1}}
\put(106,15.5){\circle{1}}
\put(108,19){\circle{1}}
\put(110,15.5){\circle{1}}
\put(121,5){\makebox(0,0){$\times 2$}}
\end{picture}

\begin{picture}(121,24)
\put(27,5){\makebox(0,0){$+$}}
\put(30,0){\line(1,0){24}}
\put(33,5){\line(1,0){18}}
\put(36,10){\line(1,0){12}}
\put(39,15){\line(1,0){6}}
\put(30,0){\line(3,5){12}}
\put(36,0){\line(3,5){9}}
\put(42,0){\line(3,5){6}}
\put(48,0){\line(3,5){3}}
\put(36,0){\line(-3,5){3}}
\put(42,0){\line(-3,5){6}}
\put(48,0){\line(-3,5){9}}
\put(54,0){\line(-3,5){12}}
\put(31,0.5){\circle*{1}}
\put(33,4){\circle*{1}}
\put(35,0.5){\circle{1}}
\put(37,0.5){\circle*{1}}
\put(39,4){\circle{1}}
\put(41,0.5){\circle*{1}}
\put(43,0.5){\circle{1}}
\put(45,4){\circle{1}}
\put(47,0.5){\circle{1}}
\put(49,0.5){\circle*{1}}
\put(51,4){\circle{1}}
\put(53,0.5){\circle*{1}}
\put(34,5.5){\circle{1}}
\put(36,9){\circle{1}}
\put(38,5.5){\circle{1}}
\put(40,5.5){\circle*{1}}
\put(42,9){\circle{1}}
\put(44,5.5){\circle*{1}}
\put(46,5.5){\circle{1}}
\put(48,9){\circle*{1}}
\put(50,5.5){\circle*{1}}
\put(37,10.5){\circle*{1}}
\put(39,14){\circle*{1}}
\put(41,10.5){\circle{1}}
\put(43,10.5){\circle*{1}}
\put(45,14){\circle*{1}}
\put(47,10.5){\circle{1}}
\put(40,15.5){\circle{1}}
\put(42,19){\circle{1}}
\put(44,15.5){\circle{1}}
\put(55,5){\makebox(0,0){$\times 2$}}
\put(60,5){\makebox(0,0){$+$}}
\put(63,0){\line(1,0){24}}
\put(66,5){\line(1,0){18}}
\put(69,10){\line(1,0){12}}
\put(72,15){\line(1,0){6}}
\put(63,0){\line(3,5){12}}
\put(69,0){\line(3,5){9}}
\put(75,0){\line(3,5){6}}
\put(81,0){\line(3,5){3}}
\put(69,0){\line(-3,5){3}}
\put(75,0){\line(-3,5){6}}
\put(81,0){\line(-3,5){9}}
\put(87,0){\line(-3,5){12}}
\put(64,0.5){\circle*{1}}
\put(66,4){\circle*{1}}
\put(68,0.5){\circle{1}}
\put(70,0.5){\circle*{1}}
\put(72,4){\circle{1}}
\put(74,0.5){\circle*{1}}
\put(76,0.5){\circle{1}}
\put(78,4){\circle{1}}
\put(80,0.5){\circle{1}}
\put(82,0.5){\circle*{1}}
\put(84,4){\circle{1}}
\put(86,0.5){\circle*{1}}
\put(67,5.5){\circle{1}}
\put(69,9){\circle{1}}
\put(71,5.5){\circle{1}}
\put(73,5.5){\circle*{1}}
\put(75,9){\circle*{1}}
\put(77,5.5){\circle{1}}
\put(79,5.5){\circle*{1}}
\put(81,9){\circle{1}}
\put(83,5.5){\circle*{1}}
\put(70,10.5){\circle*{1}}
\put(72,14){\circle*{1}}
\put(74,10.5){\circle{1}}
\put(76,10.5){\circle{1}}
\put(78,14){\circle*{1}}
\put(80,10.5){\circle*{1}}
\put(73,15.5){\circle{1}}
\put(75,19){\circle{1}}
\put(77,15.5){\circle{1}}
\put(88,5){\makebox(0,0){$\times 2$}}
\put(93,5){\makebox(0,0){$+$}}
\put(96,0){\line(1,0){24}}
\put(99,5){\line(1,0){18}}
\put(102,10){\line(1,0){12}}
\put(105,15){\line(1,0){6}}
\put(96,0){\line(3,5){12}}
\put(102,0){\line(3,5){9}}
\put(108,0){\line(3,5){6}}
\put(114,0){\line(3,5){3}}
\put(102,0){\line(-3,5){3}}
\put(108,0){\line(-3,5){6}}
\put(114,0){\line(-3,5){9}}
\put(120,0){\line(-3,5){12}}
\put(97,0.5){\circle*{1}}
\put(99,4){\circle{1}}
\put(101,0.5){\circle*{1}}
\put(103,0.5){\circle{1}}
\put(105,4){\circle{1}}
\put(107,0.5){\circle{1}}
\put(109,0.5){\circle*{1}}
\put(111,4){\circle{1}}
\put(113,0.5){\circle*{1}}
\put(115,0.5){\circle{1}}
\put(117,4){\circle*{1}}
\put(119,0.5){\circle*{1}}
\put(100,5.5){\circle*{1}}
\put(102,9){\circle*{1}}
\put(104,5.5){\circle{1}}
\put(106,5.5){\circle*{1}}
\put(108,9){\circle*{1}}
\put(110,5.5){\circle{1}}
\put(112,5.5){\circle*{1}}
\put(114,9){\circle*{1}}
\put(116,5.5){\circle{1}}
\put(103,10.5){\circle{1}}
\put(105,14){\circle{1}}
\put(107,10.5){\circle{1}}
\put(109,10.5){\circle{1}}
\put(111,14){\circle{1}}
\put(113,10.5){\circle{1}}
\put(106,15.5){\circle*{1}}
\put(108,19){\circle{1}}
\put(110,15.5){\circle*{1}}
\put(121,5){\makebox(0,0){$\times 2$}}
\end{picture}

\begin{picture}(121,24)
\put(27,5){\makebox(0,0){$+$}}
\put(30,0){\line(1,0){24}}
\put(33,5){\line(1,0){18}}
\put(36,10){\line(1,0){12}}
\put(39,15){\line(1,0){6}}
\put(30,0){\line(3,5){12}}
\put(36,0){\line(3,5){9}}
\put(42,0){\line(3,5){6}}
\put(48,0){\line(3,5){3}}
\put(36,0){\line(-3,5){3}}
\put(42,0){\line(-3,5){6}}
\put(48,0){\line(-3,5){9}}
\put(54,0){\line(-3,5){12}}
\put(31,0.5){\circle*{1}}
\put(33,4){\circle{1}}
\put(35,0.5){\circle*{1}}
\put(37,0.5){\circle{1}}
\put(39,4){\circle{1}}
\put(41,0.5){\circle{1}}
\put(43,0.5){\circle*{1}}
\put(45,4){\circle*{1}}
\put(47,0.5){\circle{1}}
\put(49,0.5){\circle*{1}}
\put(51,4){\circle{1}}
\put(53,0.5){\circle*{1}}
\put(34,5.5){\circle*{1}}
\put(36,9){\circle*{1}}
\put(38,5.5){\circle{1}}
\put(40,5.5){\circle*{1}}
\put(42,9){\circle*{1}}
\put(44,5.5){\circle{1}}
\put(46,5.5){\circle{1}}
\put(48,9){\circle*{1}}
\put(50,5.5){\circle*{1}}
\put(37,10.5){\circle{1}}
\put(39,14){\circle{1}}
\put(41,10.5){\circle{1}}
\put(43,10.5){\circle{1}}
\put(45,14){\circle{1}}
\put(47,10.5){\circle{1}}
\put(40,15.5){\circle*{1}}
\put(42,19){\circle{1}}
\put(44,15.5){\circle*{1}}
\put(55,5){\makebox(0,0){$\times 2$}}
\put(60,5){\makebox(0,0){$+$}}
\put(63,0){\line(1,0){24}}
\put(66,5){\line(1,0){18}}
\put(69,10){\line(1,0){12}}
\put(72,15){\line(1,0){6}}
\put(63,0){\line(3,5){12}}
\put(69,0){\line(3,5){9}}
\put(75,0){\line(3,5){6}}
\put(81,0){\line(3,5){3}}
\put(69,0){\line(-3,5){3}}
\put(75,0){\line(-3,5){6}}
\put(81,0){\line(-3,5){9}}
\put(87,0){\line(-3,5){12}}
\put(64,0.5){\circle*{1}}
\put(66,4){\circle{1}}
\put(68,0.5){\circle*{1}}
\put(70,0.5){\circle{1}}
\put(72,4){\circle*{1}}
\put(74,0.5){\circle*{1}}
\put(76,0.5){\circle{1}}
\put(78,4){\circle*{1}}
\put(80,0.5){\circle*{1}}
\put(82,0.5){\circle{1}}
\put(84,4){\circle*{1}}
\put(86,0.5){\circle*{1}}
\put(67,5.5){\circle*{1}}
\put(69,9){\circle*{1}}
\put(71,5.5){\circle{1}}
\put(73,5.5){\circle{1}}
\put(75,9){\circle{1}}
\put(77,5.5){\circle{1}}
\put(79,5.5){\circle{1}}
\put(81,9){\circle{1}}
\put(83,5.5){\circle{1}}
\put(70,10.5){\circle{1}}
\put(72,14){\circle{1}}
\put(74,10.5){\circle{1}}
\put(76,10.5){\circle*{1}}
\put(78,14){\circle{1}}
\put(80,10.5){\circle*{1}}
\put(73,15.5){\circle*{1}}
\put(75,19){\circle{1}}
\put(77,15.5){\circle*{1}}
\put(88,5){\makebox(0,0){$\times 2$}}
\put(93,5){\makebox(0,0){$+$}}
\put(96,0){\line(1,0){24}}
\put(99,5){\line(1,0){18}}
\put(102,10){\line(1,0){12}}
\put(105,15){\line(1,0){6}}
\put(96,0){\line(3,5){12}}
\put(102,0){\line(3,5){9}}
\put(108,0){\line(3,5){6}}
\put(114,0){\line(3,5){3}}
\put(102,0){\line(-3,5){3}}
\put(108,0){\line(-3,5){6}}
\put(114,0){\line(-3,5){9}}
\put(120,0){\line(-3,5){12}}
\put(97,0.5){\circle*{1}}
\put(99,4){\circle*{1}}
\put(101,0.5){\circle{1}}
\put(103,0.5){\circle*{1}}
\put(105,4){\circle{1}}
\put(107,0.5){\circle*{1}}
\put(109,0.5){\circle{1}}
\put(111,4){\circle*{1}}
\put(113,0.5){\circle*{1}}
\put(115,0.5){\circle{1}}
\put(117,4){\circle*{1}}
\put(119,0.5){\circle*{1}}
\put(100,5.5){\circle{1}}
\put(102,9){\circle*{1}}
\put(104,5.5){\circle*{1}}
\put(106,5.5){\circle{1}}
\put(108,9){\circle{1}}
\put(110,5.5){\circle{1}}
\put(112,5.5){\circle{1}}
\put(114,9){\circle{1}}
\put(116,5.5){\circle{1}}
\put(103,10.5){\circle{1}}
\put(105,14){\circle{1}}
\put(107,10.5){\circle{1}}
\put(109,10.5){\circle*{1}}
\put(111,14){\circle{1}}
\put(113,10.5){\circle*{1}}
\put(106,15.5){\circle*{1}}
\put(108,19){\circle{1}}
\put(110,15.5){\circle*{1}}
\put(121,5){\makebox(0,0){$\times 2$}}
\end{picture}

\begin{picture}(121,24)
\put(27,5){\makebox(0,0){$+$}}
\put(30,0){\line(1,0){24}}
\put(33,5){\line(1,0){18}}
\put(36,10){\line(1,0){12}}
\put(39,15){\line(1,0){6}}
\put(30,0){\line(3,5){12}}
\put(36,0){\line(3,5){9}}
\put(42,0){\line(3,5){6}}
\put(48,0){\line(3,5){3}}
\put(36,0){\line(-3,5){3}}
\put(42,0){\line(-3,5){6}}
\put(48,0){\line(-3,5){9}}
\put(54,0){\line(-3,5){12}}
\put(31,0.5){\circle*{1}}
\put(33,4){\circle{1}}
\put(35,0.5){\circle*{1}}
\put(37,0.5){\circle{1}}
\put(39,4){\circle*{1}}
\put(41,0.5){\circle*{1}}
\put(43,0.5){\circle{1}}
\put(45,4){\circle{1}}
\put(47,0.5){\circle{1}}
\put(49,0.5){\circle*{1}}
\put(51,4){\circle{1}}
\put(53,0.5){\circle*{1}}
\put(34,5.5){\circle*{1}}
\put(36,9){\circle*{1}}
\put(38,5.5){\circle{1}}
\put(40,5.5){\circle{1}}
\put(42,9){\circle{1}}
\put(44,5.5){\circle{1}}
\put(46,5.5){\circle*{1}}
\put(48,9){\circle{1}}
\put(50,5.5){\circle*{1}}
\put(37,10.5){\circle{1}}
\put(39,14){\circle{1}}
\put(41,10.5){\circle{1}}
\put(43,10.5){\circle*{1}}
\put(45,14){\circle{1}}
\put(47,10.5){\circle*{1}}
\put(40,15.5){\circle*{1}}
\put(42,19){\circle{1}}
\put(44,15.5){\circle*{1}}
\put(55,5){\makebox(0,0){$\times 2$}}
\put(60,5){\makebox(0,0){$+$}}
\put(63,0){\line(1,0){24}}
\put(66,5){\line(1,0){18}}
\put(69,10){\line(1,0){12}}
\put(72,15){\line(1,0){6}}
\put(63,0){\line(3,5){12}}
\put(69,0){\line(3,5){9}}
\put(75,0){\line(3,5){6}}
\put(81,0){\line(3,5){3}}
\put(69,0){\line(-3,5){3}}
\put(75,0){\line(-3,5){6}}
\put(81,0){\line(-3,5){9}}
\put(87,0){\line(-3,5){12}}
\put(64,0.5){\circle*{1}}
\put(66,4){\circle*{1}}
\put(68,0.5){\circle{1}}
\put(70,0.5){\circle*{1}}
\put(72,4){\circle{1}}
\put(74,0.5){\circle*{1}}
\put(76,0.5){\circle{1}}
\put(78,4){\circle{1}}
\put(80,0.5){\circle{1}}
\put(82,0.5){\circle*{1}}
\put(84,4){\circle{1}}
\put(86,0.5){\circle*{1}}
\put(67,5.5){\circle{1}}
\put(69,9){\circle*{1}}
\put(71,5.5){\circle*{1}}
\put(73,5.5){\circle{1}}
\put(75,9){\circle{1}}
\put(77,5.5){\circle{1}}
\put(79,5.5){\circle*{1}}
\put(81,9){\circle{1}}
\put(83,5.5){\circle*{1}}
\put(70,10.5){\circle{1}}
\put(72,14){\circle{1}}
\put(74,10.5){\circle{1}}
\put(76,10.5){\circle*{1}}
\put(78,14){\circle{1}}
\put(80,10.5){\circle*{1}}
\put(73,15.5){\circle*{1}}
\put(75,19){\circle{1}}
\put(77,15.5){\circle*{1}}
\put(88,5){\makebox(0,0){$\times 2$}}
\end{picture}

\caption{\footnotesize{Illustration for the expression of $h_{2,4}(n+1)$ with non-negative $n$. The multiplication of two on the right-hand-side corresponds to the reflection symmetry with respect to the central vertical axis.}} 
\label{h24fig}
\end{figure}
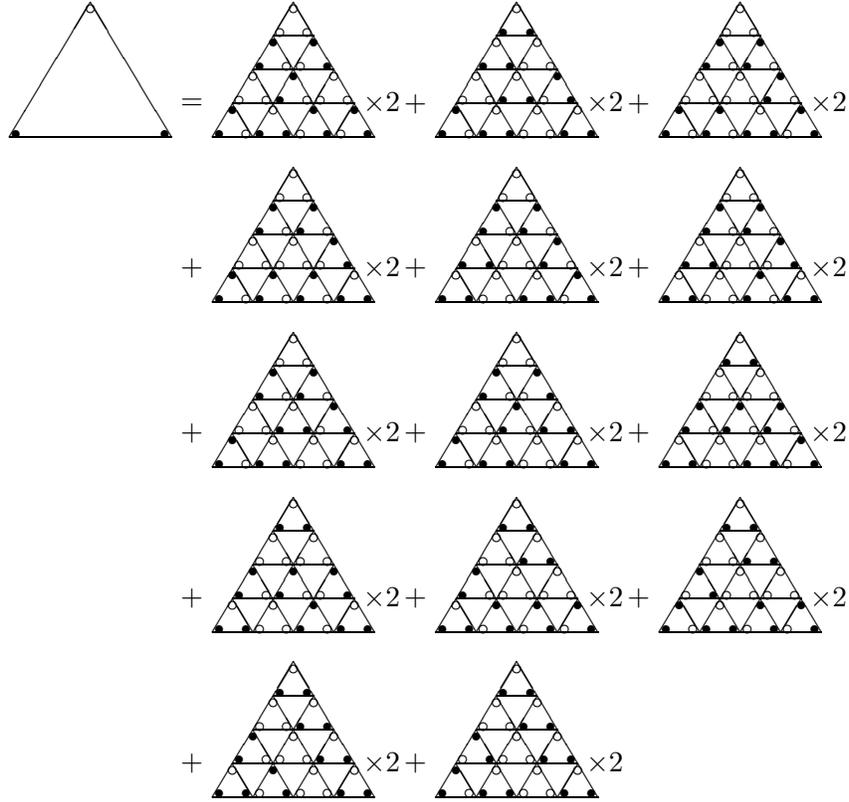

Similarly, $h_{2,4}(n+1)$ with non-negative $n$ for $SG_{2,4}(n+1)$ can be obtained with appropriate configurations of its ten constituting $SG_{2,4}(n)$ as illustrated in Fig. \ref{h24fig} to verify Eq. (\ref{h24eq}). \ $\Box$

\bigskip

It is elementary to solve $f_{2,4}(n)$ and $h_{2,4}(n)$ in order to obtain the entropy for $SG_{2,4}$.

\bigskip

\begin{theo} \label{theosg24} For the generalized two-dimensional Sierpinski gasket $SG_{2,4}(n)$ with non-negative integer $n$,
\beq
f_{2,4}(n) = h_{2,4}(n) = 28^{\gamma_{2,4} (n)} \ ,
\label{fh24}
\eeq
where the exponent is
\beq
\gamma_{2,4} (n) = \frac19 (10^n-1) \ .
\label{alpha24}
\eeq
Define the number of dimer coverings $N(SG_{2,4}(n))$ in Eq. (\ref{sdef}) equal to $f_{2,4}(n)$. With $v(SG_{2,4}(n))=(4(10)^n+5)/3$, the entropy is given by 
\beq
S_{SG_{2,4}} = \frac{1}{12} \ln 28 \simeq 0.27768370918...
\label{Ssg24}
\eeq
\end{theo}

\bigskip

\subsection{$SG_{2,5}(n)$}

For the generalized two-dimensional Sierpinski gasket $SG_{2,b}(n)$ with $b=5$, we have
\beq
v(SG_{2,5}(n)) = \frac{9(15)^n+12}{7} = 15^n + 2 + \frac27 (15^n-1) 
= 15^n + 2 + \frac27 \sum_{j=1}^n {n \choose j} 14^j
\eeq
by Eq. (\ref{bv}), such that the number of vertices is always odd for any $n$. Therefore, $g_{2,5}(n)$ and $t_{2,5}(n)$ are zero for all $n$, while the initial values remain $f_{2,5}(0)=1$, $g_{2,5}(0)=0$, $h_{2,5}(0)=1$ and $t_{2,5}(0)=0$. The figures of the recursion relations for $f_{2,5}(n)$ and $h_{2,5}(n)$ are too many to be shown here, and we state the following Lemma without proof.

\bigskip

\begin{lemma} \label{lemmasg25r} For any non-negative integer $n$,
\beq
f_{2,5}(n+1) = 200 f_{2,5}^6(n) h_{2,5}^9(n) \ , 
\label{f25eq}
\eeq
\beq
h_{2,5}(n+1) = 200 f_{2,5}^5(n) h_{2,5}^{10}(n) \ .
\label{h25eq}
\eeq
\end{lemma}

\bigskip

It is elementary to solve $f_{2,5}(n)$ and $h_{2,5}(n)$ in order to obtain the entropy for $SG_{2,5}$.

\bigskip

\begin{theo} \label{theosg25} For the generalized two-dimensional Sierpinski gasket $SG_{2,5}(n)$ with non-negative integer $n$,
\beq
f_{2,5}(n) = h_{2,5}(n) = 200^{\gamma_{2,5} (n)} \ ,
\label{fh25}
\eeq
where the exponent is
\beq
\gamma_{2,5} (n) = \frac{1}{14} (15^n-1) \ .
\label{alpha25}
\eeq
Define the number of dimer coverings $N(SG_{2,5}(n))$ in Eq. (\ref{sdef}) equal to $f_{2,5}(n)$. With $v(SG_{2,5}(n))=(9(15)^n+12)/7$, the entropy is given by 
\beq
S_{SG_{2,5}} = \frac{1}{18} \ln 200 \simeq 0.29435096480...
\label{Ssg25}
\eeq
\end{theo}

As the generalized two-dimensional Sierpinski gasket $SG_{2,b}(n)$ for any $b$ is planar, it appears that the number of dimer coverings can be solved exactly. However, the number of configurations to be considered increases as $b$ increases and the recursion relations must be derived individually for each $b$. We have been unable to obtain a general expression of the number of dimer coverings on $SG_{2,b}(n)$ for arbitrary $b$.

\section{The number of dimer coverings on $SG_d(n)$ with $d=3,4,5$} 
\label{sectionIV}

In this section we present the number of dimer coverings on the Sierpinski gasket $SG_d(n)$ with $d=3,4,5$ which is not planar. Instead of solving exactly the entropies for these Sierpinski gaskets, we obtain accurate upper and lower bounds for them.

\subsection{$SG_3(n)$}

For the three-dimensional Sierpinski gasket $SG_3(n)$, we use the following definitions.

\bigskip

\begin{defi} \label{defisg3} Consider the three-dimensional Sierpinski gasket $SG_3(n)$ at stage $n$. (i) Define $f_3(n)$ as the number of dimer coverings such that the four outmost vertices are vacant. (ii) Define $h_3(n)$ as the number of dimer coverings such that two certain outmost vertices are occupied by dimers and the other two outmost vertices are vacant. (iii) Define $s_3(n)$ as the number of dimer coverings such that all four outmost vertices are occupied by dimers.
\end{defi}

\bigskip

\begin{figure}[htbp]
\unitlength 1.8mm 
\begin{picture}(30,5)
\put(0,0){\line(1,0){6}}
\put(0,0){\line(3,5){3}}
\put(6,0){\line(-3,5){3}}
\put(0,0){\line(3,2){3}}
\put(6,0){\line(-3,2){3}}
\put(3,2){\line(0,1){3}}
\put(1,0.5){\circle{1}}
\put(5,0.5){\circle{1}}
\put(3,2){\circle{1}}
\put(3,4){\circle{1}}
\put(3,-2){\makebox(0,0){$f_3(n)$}}
\put(12,0){\line(1,0){6}}
\put(12,0){\line(3,5){3}}
\put(18,0){\line(-3,5){3}}
\put(12,0){\line(3,2){3}}
\put(18,0){\line(-3,2){3}}
\put(15,2){\line(0,1){3}}
\put(13,0.5){\circle*{1}}
\put(17,0.5){\circle*{1}}
\put(15,2){\circle{1}}
\put(15,4){\circle{1}}
\put(15,-2){\makebox(0,0){$h_3(n)$}}
\put(24,0){\line(1,0){6}}
\put(24,0){\line(3,5){3}}
\put(30,0){\line(-3,5){3}}
\put(24,0){\line(3,2){3}}
\put(30,0){\line(-3,2){3}}
\put(27,2){\line(0,1){3}}
\put(25,0.5){\circle*{1}}
\put(29,0.5){\circle*{1}}
\put(27,2){\circle*{1}}
\put(27,4){\circle*{1}}
\put(27,-2){\makebox(0,0){$s_3(n)$}}
\end{picture}

\vspace*{5mm}
\caption{\footnotesize{Illustration for the dimer coverings $f_3(n)$, $h_3(n)$ and $s_3(n)$. Only the four outmost vertices are shown explicitly, where each open circle is vacant and each solid circle is occupied by a dimer.}} 
\label{fhsfig}
\end{figure}
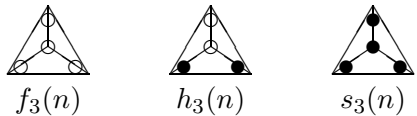

As the number of vertices for $SG_3(n)$ is always even by Eq. (\ref{v}), we do not have the dimer coverings such that one certain outmost vertices is occupied by a dimer and the other three outmost vertices are vacant, or one certain outmost vertices is vacant and the other three outmost vertices are occupied by dimers.
The quantities $f_3(n)$, $h_3(n)$, and $s_3(n)$ are illustrated in Fig. \ref{fhsfig}, where only the outmost vertices are shown. There are ${4 \choose 2}=6$ equivalent $h_3(n)$. The initial values at stage zero are $f_3(0)=1$, $h_3(0)=1$, $s_3(0)=3$. These quantities satisfy recursion relations.

\bigskip

\bigskip

\begin{lemma} \label{lemmasg3r} For any non-negative integer $n$,
\beq
f_3(n+1) = 8 f_3(n) h_3^3(n) \ , 
\label{f3eq}
\eeq
\beq
h_3(n+1) = 4 f_3(n) h_3^2(n) s_3(n) + 4 h_3^4(n) \ , 
\label{h3eq}
\eeq
\beq
s_3(n+1) = 8 h_3^3(n) s_3(n) \ . 
\label{s3eq}
\eeq
\end{lemma}

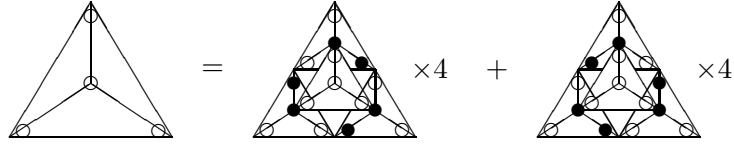
\begin{figure}[htbp]
\unitlength 1.8mm 
\begin{picture}(52,12)
\put(0,0){\line(1,0){12}}
\put(0,0){\line(3,5){6}}
\put(12,0){\line(-3,5){6}}
\put(0,0){\line(3,2){6}}
\put(12,0){\line(-3,2){6}}
\put(6,4){\line(0,1){6}}
\put(6,4){\circle{1}}
\multiput(1,0.5)(10,0){2}{\circle{1}}
\put(6,9){\circle{1}}
\put(15,5){\makebox(0,0){$=$}}
\put(18,0){\line(1,0){12}}
\put(21,5){\line(1,0){1.8}}
\put(27,5){\line(-1,0){1.8}}
\put(18,0){\line(3,5){6}}
\put(24,0){\line(3,5){1.2}}
\put(27,5){\line(-3,-5){1.2}}
\put(24,0){\line(-3,5){1.2}}
\put(21,5){\line(3,-5){1.2}}
\put(30,0){\line(-3,5){6}}
\multiput(18,0)(6,0){2}{\line(3,2){3}}
\put(21,5){\line(3,2){3}}
\multiput(30,0)(-6,0){2}{\line(-3,2){3}}
\put(27,5){\line(-3,2){3}}
\multiput(21,2)(6,0){2}{\line(0,1){3}}
\put(24,7){\line(0,1){3}}
\put(21,2){\line(1,0){6}}
\put(21,2){\line(3,5){3}}
\put(27,2){\line(-3,5){3}}
\put(21,2){\line(3,2){3}}
\put(27,2){\line(-3,2){3}}
\put(24,4){\line(0,1){3}}
\put(19,0.5){\circle{1}}
\put(21,4){\circle*{1}}
\put(23,0.5){\circle{1}}
\put(21,2){\circle*{1}}
\put(22,5.5){\circle{1}}
\put(24,9){\circle{1}}
\put(26,5.5){\circle*{1}}
\put(24,7){\circle*{1}}
\put(25,0.5){\circle*{1}}
\put(27,4){\circle{1}}
\put(29,0.5){\circle{1}}
\put(27,2){\circle*{1}}
\put(22,2.5){\circle{1}}
\put(24,6){\circle{1}}
\put(26,2.5){\circle{1}}
\put(24,4){\circle{1}}
\put(31,5){\makebox(0,0){$\times 4$}}
\put(36,5){\makebox(0,0){$+$}}
\put(39,0){\line(1,0){12}}
\put(42,5){\line(1,0){1.8}}
\put(48,5){\line(-1,0){1.8}}
\put(39,0){\line(3,5){6}}
\put(45,0){\line(3,5){1.2}}
\put(48,5){\line(-3,-5){1.2}}
\put(45,0){\line(-3,5){1.2}}
\put(42,5){\line(3,-5){1.2}}
\put(51,0){\line(-3,5){6}}
\multiput(39,0)(6,0){2}{\line(3,2){3}}
\put(42,5){\line(3,2){3}}
\multiput(51,0)(-6,0){2}{\line(-3,2){3}}
\put(48,5){\line(-3,2){3}}
\multiput(42,2)(6,0){2}{\line(0,1){3}}
\put(45,7){\line(0,1){3}}
\put(42,2){\line(1,0){6}}
\put(42,2){\line(3,5){3}}
\put(48,2){\line(-3,5){3}}
\put(42,2){\line(3,2){3}}
\put(48,2){\line(-3,2){3}}
\put(45,4){\line(0,1){3}}
\put(40,0.5){\circle{1}}
\put(42,4){\circle{1}}
\put(44,0.5){\circle*{1}}
\put(42,2){\circle*{1}}
\put(43,5.5){\circle*{1}}
\put(45,9){\circle{1}}
\put(47,5.5){\circle{1}}
\put(45,7){\circle*{1}}
\put(46,0.5){\circle{1}}
\put(48,4){\circle*{1}}
\put(50,0.5){\circle{1}}
\put(48,2){\circle*{1}}
\put(43,2.5){\circle{1}}
\put(45,6){\circle{1}}
\put(47,2.5){\circle{1}}
\put(45,4){\circle{1}}
\put(52,5){\makebox(0,0){$\times 4$}}
\end{picture}

\caption{\footnotesize{Illustration for the expression of $f_3(n+1)$. The multiplication of four on the right-hand-side corresponds to the four possible orientations of $SG_3(n+1)$.}} 
\label{f3fig}
\end{figure}

{\sl Proof} \quad 
The Sierpinski gasket $SG_3(n+1)$ is composed of four $SG_3(n)$ with six pairs of vertices identified. For each pair of identified vertices, either one of them is originally occupied by a dimer while the other one is vacant. The number $f_3(n+1)$ for non-negative $n$ consists of eight configurations where one of the $SG_3(n)$ are in the $f_3(n)$ status and the other three are in the $h_3(n)$ status as illustrated in Fig. \ref{f3fig}, such that Eq. (\ref{f3eq}) is verified. 

\bigskip

\bigskip

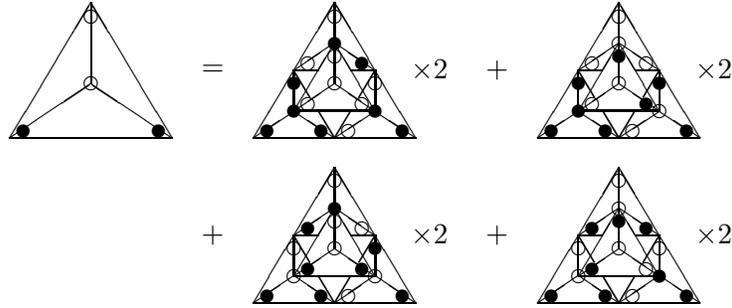
\begin{figure}[htbp]
\unitlength 1.8mm 
\begin{picture}(52,12)
\put(0,0){\line(1,0){12}}
\put(0,0){\line(3,5){6}}
\put(12,0){\line(-3,5){6}}
\put(0,0){\line(3,2){6}}
\put(12,0){\line(-3,2){6}}
\put(6,4){\line(0,1){6}}
\put(6,4){\circle{1}}
\multiput(1,0.5)(10,0){2}{\circle*{1}}
\put(6,9){\circle{1}}
\put(15,5){\makebox(0,0){$=$}}
\put(18,0){\line(1,0){12}}
\put(21,5){\line(1,0){1.8}}
\put(27,5){\line(-1,0){1.8}}
\put(18,0){\line(3,5){6}}
\put(24,0){\line(3,5){1.2}}
\put(27,5){\line(-3,-5){1.2}}
\put(24,0){\line(-3,5){1.2}}
\put(21,5){\line(3,-5){1.2}}
\put(30,0){\line(-3,5){6}}
\multiput(18,0)(6,0){2}{\line(3,2){3}}
\put(21,5){\line(3,2){3}}
\multiput(30,0)(-6,0){2}{\line(-3,2){3}}
\put(27,5){\line(-3,2){3}}
\multiput(21,2)(6,0){2}{\line(0,1){3}}
\put(24,7){\line(0,1){3}}
\put(21,2){\line(1,0){6}}
\put(21,2){\line(3,5){3}}
\put(27,2){\line(-3,5){3}}
\put(21,2){\line(3,2){3}}
\put(27,2){\line(-3,2){3}}
\put(24,4){\line(0,1){3}}
\put(19,0.5){\circle*{1}}
\put(21,4){\circle*{1}}
\put(23,0.5){\circle*{1}}
\put(21,2){\circle*{1}}
\put(22,5.5){\circle{1}}
\put(24,9){\circle{1}}
\put(26,5.5){\circle*{1}}
\put(24,7){\circle*{1}}
\put(25,0.5){\circle{1}}
\put(27,4){\circle{1}}
\put(29,0.5){\circle*{1}}
\put(27,2){\circle*{1}}
\put(22,2.5){\circle{1}}
\put(24,6){\circle{1}}
\put(26,2.5){\circle{1}}
\put(24,4){\circle{1}}
\put(31,5){\makebox(0,0){$\times 2$}}
\put(36,5){\makebox(0,0){$+$}}
\put(39,0){\line(1,0){12}}
\put(42,5){\line(1,0){1.8}}
\put(48,5){\line(-1,0){1.8}}
\put(39,0){\line(3,5){6}}
\put(45,0){\line(3,5){1.2}}
\put(48,5){\line(-3,-5){1.2}}
\put(45,0){\line(-3,5){1.2}}
\put(42,5){\line(3,-5){1.2}}
\put(51,0){\line(-3,5){6}}
\multiput(39,0)(6,0){2}{\line(3,2){3}}
\put(42,5){\line(3,2){3}}
\multiput(51,0)(-6,0){2}{\line(-3,2){3}}
\put(48,5){\line(-3,2){3}}
\multiput(42,2)(6,0){2}{\line(0,1){3}}
\put(45,7){\line(0,1){3}}
\put(42,2){\line(1,0){6}}
\put(42,2){\line(3,5){3}}
\put(48,2){\line(-3,5){3}}
\put(42,2){\line(3,2){3}}
\put(48,2){\line(-3,2){3}}
\put(45,4){\line(0,1){3}}
\put(40,0.5){\circle*{1}}
\put(42,4){\circle*{1}}
\put(44,0.5){\circle*{1}}
\put(42,2){\circle*{1}}
\put(43,5.5){\circle{1}}
\put(45,9){\circle{1}}
\put(47,5.5){\circle{1}}
\put(45,7){\circle{1}}
\put(46,0.5){\circle{1}}
\put(48,4){\circle*{1}}
\put(50,0.5){\circle*{1}}
\put(48,2){\circle{1}}
\put(43,2.5){\circle{1}}
\put(45,6){\circle*{1}}
\put(47,2.5){\circle*{1}}
\put(45,4){\circle{1}}
\put(52,5){\makebox(0,0){$\times 2$}}
\end{picture}

\begin{picture}(52,12)
\put(15,5){\makebox(0,0){$+$}}
\put(18,0){\line(1,0){12}}
\put(21,5){\line(1,0){1.8}}
\put(27,5){\line(-1,0){1.8}}
\put(18,0){\line(3,5){6}}
\put(24,0){\line(3,5){1.2}}
\put(27,5){\line(-3,-5){1.2}}
\put(24,0){\line(-3,5){1.2}}
\put(21,5){\line(3,-5){1.2}}
\put(30,0){\line(-3,5){6}}
\multiput(18,0)(6,0){2}{\line(3,2){3}}
\put(21,5){\line(3,2){3}}
\multiput(30,0)(-6,0){2}{\line(-3,2){3}}
\put(27,5){\line(-3,2){3}}
\multiput(21,2)(6,0){2}{\line(0,1){3}}
\put(24,7){\line(0,1){3}}
\put(21,2){\line(1,0){6}}
\put(21,2){\line(3,5){3}}
\put(27,2){\line(-3,5){3}}
\put(21,2){\line(3,2){3}}
\put(27,2){\line(-3,2){3}}
\put(24,4){\line(0,1){3}}
\put(19,0.5){\circle*{1}}
\put(21,4){\circle{1}}
\put(23,0.5){\circle*{1}}
\put(21,2){\circle{1}}
\put(22,5.5){\circle*{1}}
\put(24,9){\circle{1}}
\put(26,5.5){\circle{1}}
\put(24,7){\circle*{1}}
\put(25,0.5){\circle{1}}
\put(27,4){\circle*{1}}
\put(29,0.5){\circle*{1}}
\put(27,2){\circle{1}}
\put(22,2.5){\circle*{1}}
\put(24,6){\circle{1}}
\put(26,2.5){\circle*{1}}
\put(24,4){\circle{1}}
\put(31,5){\makebox(0,0){$\times 2$}}
\put(36,5){\makebox(0,0){$+$}}
\put(39,0){\line(1,0){12}}
\put(42,5){\line(1,0){1.8}}
\put(48,5){\line(-1,0){1.8}}
\put(39,0){\line(3,5){6}}
\put(45,0){\line(3,5){1.2}}
\put(48,5){\line(-3,-5){1.2}}
\put(45,0){\line(-3,5){1.2}}
\put(42,5){\line(3,-5){1.2}}
\put(51,0){\line(-3,5){6}}
\multiput(39,0)(6,0){2}{\line(3,2){3}}
\put(42,5){\line(3,2){3}}
\multiput(51,0)(-6,0){2}{\line(-3,2){3}}
\put(48,5){\line(-3,2){3}}
\multiput(42,2)(6,0){2}{\line(0,1){3}}
\put(45,7){\line(0,1){3}}
\put(42,2){\line(1,0){6}}
\put(42,2){\line(3,5){3}}
\put(48,2){\line(-3,5){3}}
\put(42,2){\line(3,2){3}}
\put(48,2){\line(-3,2){3}}
\put(45,4){\line(0,1){3}}
\put(40,0.5){\circle*{1}}
\put(42,4){\circle{1}}
\put(44,0.5){\circle*{1}}
\put(42,2){\circle{1}}
\put(43,5.5){\circle*{1}}
\put(45,9){\circle{1}}
\put(47,5.5){\circle*{1}}
\put(45,7){\circle{1}}
\put(46,0.5){\circle{1}}
\put(48,4){\circle{1}}
\put(50,0.5){\circle*{1}}
\put(48,2){\circle*{1}}
\put(43,2.5){\circle*{1}}
\put(45,6){\circle*{1}}
\put(47,2.5){\circle{1}}
\put(45,4){\circle{1}}
\put(52,5){\makebox(0,0){$\times 2$}}
\end{picture}

\caption{\footnotesize{Illustration for the expression of $h_3(n+1)$. The multiplication of two on the right-hand-side corresponds to the reflection symmetry with respect to the central vertical axis.}} 
\label{h3fig}
\end{figure}

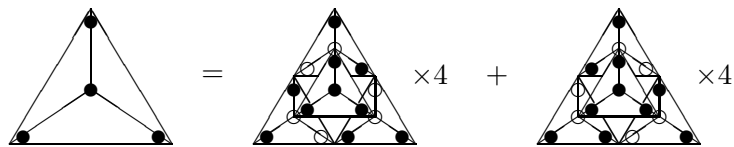
\begin{figure}[htbp]
\unitlength 1.8mm 
\begin{picture}(52,12)
\put(0,0){\line(1,0){12}}
\put(0,0){\line(3,5){6}}
\put(12,0){\line(-3,5){6}}
\put(0,0){\line(3,2){6}}
\put(12,0){\line(-3,2){6}}
\put(6,4){\line(0,1){6}}
\put(6,4){\circle*{1}}
\multiput(1,0.5)(10,0){2}{\circle*{1}}
\put(6,9){\circle*{1}}
\put(15,5){\makebox(0,0){$=$}}
\put(18,0){\line(1,0){12}}
\put(21,5){\line(1,0){1.8}}
\put(27,5){\line(-1,0){1.8}}
\put(18,0){\line(3,5){6}}
\put(24,0){\line(3,5){1.2}}
\put(27,5){\line(-3,-5){1.2}}
\put(24,0){\line(-3,5){1.2}}
\put(21,5){\line(3,-5){1.2}}
\put(30,0){\line(-3,5){6}}
\multiput(18,0)(6,0){2}{\line(3,2){3}}
\put(21,5){\line(3,2){3}}
\multiput(30,0)(-6,0){2}{\line(-3,2){3}}
\put(27,5){\line(-3,2){3}}
\multiput(21,2)(6,0){2}{\line(0,1){3}}
\put(24,7){\line(0,1){3}}
\put(21,2){\line(1,0){6}}
\put(21,2){\line(3,5){3}}
\put(27,2){\line(-3,5){3}}
\put(21,2){\line(3,2){3}}
\put(27,2){\line(-3,2){3}}
\put(24,4){\line(0,1){3}}
\put(19,0.5){\circle*{1}}
\put(21,4){\circle*{1}}
\put(23,0.5){\circle{1}}
\put(21,2){\circle{1}}
\put(22,5.5){\circle{1}}
\put(24,9){\circle*{1}}
\put(26,5.5){\circle*{1}}
\put(24,7){\circle{1}}
\put(25,0.5){\circle*{1}}
\put(27,4){\circle{1}}
\put(29,0.5){\circle*{1}}
\put(27,2){\circle{1}}
\put(22,2.5){\circle*{1}}
\put(24,6){\circle*{1}}
\put(26,2.5){\circle*{1}}
\put(24,4){\circle*{1}}
\put(31,5){\makebox(0,0){$\times 4$}}
\put(36,5){\makebox(0,0){$+$}}
\put(39,0){\line(1,0){12}}
\put(42,5){\line(1,0){1.8}}
\put(48,5){\line(-1,0){1.8}}
\put(39,0){\line(3,5){6}}
\put(45,0){\line(3,5){1.2}}
\put(48,5){\line(-3,-5){1.2}}
\put(45,0){\line(-3,5){1.2}}
\put(42,5){\line(3,-5){1.2}}
\put(51,0){\line(-3,5){6}}
\multiput(39,0)(6,0){2}{\line(3,2){3}}
\put(42,5){\line(3,2){3}}
\multiput(51,0)(-6,0){2}{\line(-3,2){3}}
\put(48,5){\line(-3,2){3}}
\multiput(42,2)(6,0){2}{\line(0,1){3}}
\put(45,7){\line(0,1){3}}
\put(42,2){\line(1,0){6}}
\put(42,2){\line(3,5){3}}
\put(48,2){\line(-3,5){3}}
\put(42,2){\line(3,2){3}}
\put(48,2){\line(-3,2){3}}
\put(45,4){\line(0,1){3}}
\put(40,0.5){\circle*{1}}
\put(42,4){\circle{1}}
\put(44,0.5){\circle*{1}}
\put(42,2){\circle{1}}
\put(43,5.5){\circle*{1}}
\put(45,9){\circle*{1}}
\put(47,5.5){\circle{1}}
\put(45,7){\circle{1}}
\put(46,0.5){\circle{1}}
\put(48,4){\circle*{1}}
\put(50,0.5){\circle*{1}}
\put(48,2){\circle{1}}
\put(43,2.5){\circle*{1}}
\put(45,6){\circle*{1}}
\put(47,2.5){\circle*{1}}
\put(45,4){\circle*{1}}
\put(52,5){\makebox(0,0){$\times 4$}}
\end{picture}

\caption{\footnotesize{Illustration for the expression of $s_3(n+1)$. The multiplication of four on the right-hand-side corresponds to the four possible orientations of $SG_3(n+1)$.}} 
\label{s3fig}
\end{figure}

Similarly, $h_3(n+1)$ and $s_3(n+1)$ for $SG_3(n+1)$ can be obtained with appropriate configurations of its four constituting $SG_3(n)$ as illustrated in Figs. \ref{h3fig}, and \ref{s3fig} to verify Eqs. (\ref{h3eq}) and (\ref{s3eq}), respectively. 
\ $\Box$

\bigskip

\bigskip

\bigskip

The values of $f_3(n)$, $h_3(n)$, $s_3(n)$ for small $n$ can be evaluated recursively by Eqs. (\ref{f3eq})-(\ref{s3eq}), but they grow exponentially, and do not have simple integer factorizations. To estimate the value of entropy for $SG_3$, we define the ratio
\beq
\alpha_3(n) = \frac{h_3(n)}{f_3(n)} \ ,
\label{alpha3def}
\eeq
and its limit
\beq
\alpha_3 \equiv \lim_{n \to \infty} \alpha_3(n) \ .
\eeq

\bigskip

\begin{lemma} \label{lemmasg3d} Sequence $\{ \alpha_3(n) \}_{n=1}^{\infty}$ decreases monotonically. The limit $\alpha_3$ is equal to $\sqrt{3}$.
\end{lemma}

{\sl Proof} \quad From Eqs. (\ref{f3eq}) and (\ref{s3eq}), the ratio $s_3(n)/f_3(n)$ is invariant, that is equal to $s_3(0)/f_3(0)=3$. Eq. (\ref{h3eq}) can be modified to be
\beq
h_3(n+1) = 12 f_3^2(n) h_3^2(n) + 4 h_3^4(n) \ .
\label{h3eqn}
\eeq
Although $\alpha_3(0)=1$, it is clear that $\alpha_3(n)$ is bounded below by $\sqrt{3}$ for positive integer $n$ because
\beq
h_3^2(n+1)-3f_3^2(n+1) = [12f_3^2(n)h_3^2(n) - 4h_3^4(n)]^2 \ge 0 \ .
\eeq
It follows that $\alpha_3(n)$ decreases for positive $n$ because
\beq
\frac{h_3(n)}{f_3(n)} - \frac{h_3(n+1)}{f_3(n+1)} = \frac{h_3^2(n)-3f_3^2(n)}{2f_3(n)h_3(n)} \ge 0 \ ,
\eeq
which implies that the limit $\alpha_3$ exists. From Eqs. (\ref{f3eq}) and (\ref{h3eqn}), we have
\beq
\frac{h_3(n+1)}{f_3(n+1)} = \frac32 \frac{f_3(n)}{h_3(n)} + \frac12 \frac{h_3(n)}{f_3(n)} \ .
\label{h3overf3}
\eeq
By taking the large $n$ limit in Eq. (\ref{h3overf3}), $\alpha_3$ is solved to be $\sqrt{3}$.
\ $\Box$

\bigskip

The general expressions for $f_3(n)$ and $h_3(n)$ in terms of quantities at stage $m<n$ can be written as follows.

\bigskip

\begin{lemma} \label{lemmasg3fh} For a non-negative integer $m$ and any positive integer $n > m$, 
\beqs
f_3(n) & = & 2^{\frac{2(4)^{n-m+1}-5-3(-1)^{n-m}}{10}} f_3(m)^{\frac{2(4)^{n-m}+3(-1)^{n-m}}{5}} 
h_3(m)^{\frac{3(4)^{n-m}-3(-1)^{n-m}}{5}} \cr\cr
& & \times \prod _{j=2}^{n-m} \Big [ 3 + \alpha_3^2(n-j) \Big ]^{\frac{3(4)^{j-1}-3(-1)^{j-1}}{5}} \ , \cr & &
\label{f3gen}
\eeqs
\beqs
h_3(n) & = & 2^{\frac{4^{n-m+1}-5+(-1)^{n-m}}{5}} f_3(m)^{\frac{2(4)^{n-m}-2(-1)^{n-m}}{5}} h_3(m)^{\frac{3(4)^{n-m}+2(-1)^{n-m}}{5}} \cr\cr & & 
\times \prod _{j=1}^{n-m} \Big [ 3 + \alpha_3^2(n-j) \Big ]^{\frac{3(4)^{j-1}+2(-1)^{j-1}}{5}} \ .
\label{h3gen}
\eeqs
Here when $n-m=1$, the product with lower limit two is defined to be one.
\end{lemma}

{\sl Proof} \quad 
It is clear that Eqs. (\ref{f3gen}) and (\ref{h3gen}) are valid for $n=m+1$ since $f_3(m+1)=8f_3(m)h_3^3(m)$ and $h_3(m+1)=4f_3^2(m)h_3^2(m)[3+\alpha_3^2(m)]$ by Eqs. (\ref{f3eq}) and (\ref{h3eqn}), respectively. Consider Eq. (\ref{f3gen}) holds for a certain positive integer $n>m$, then 
\beqs
\lefteqn{f_3(n+1) = 8 f_3(n) h_3^3(n)} \cr\cr
& = & 8 \times 2^{\frac{2(4)^{n-m+1}-5-3(-1)^{n-m}}{10}} f_3(m)^{\frac{2(4)^{n-m}+3(-1)^{n-m}}{5}} h_3(m)^{\frac{3(4)^{n-m}-3(-1)^{n-m}}{5}}  \cr\cr 
& & \times 
\prod _{j=2}^{n-m} \Big [ 3 + \alpha_3^2(n-j) \Big ]^{\frac{3(4)^{j-1}-3(-1)^{j-1}}{5}} 2^{\frac{3(4)^{n-m+1}-15+3(-1)^{n-m}}{5}} f_3(m)^{\frac{6(4)^{n-m}-6(-1)^{n-m}}{5}} \cr\cr 
& & \times h_3(m)^{\frac{9(4)^{n-m}+6(-1)^{n-m}}{5}} \prod _{j=1}^{n-m} \Big [ 3 + \alpha_3^2(n-j) \Big ]^{\frac{9(4)^{j-1}+6(-1)^{j-1}}{5}} \cr\cr
& = & 2^{\frac{8(4)^{n-m+1}-5+3(-1)^{n-m}}{10}} f_3(m)^{\frac{8(4)^{n-m}-3(-1)^{n-m}}{5}} h_3(m)^{\frac{12(4)^{n-m}+3(-1)^{n-m}}{5}} \cr\cr 
& & \times \prod _{j=2}^{n-m} \Big [ 3 + \alpha_3^2(n-j) \Big ]^{\frac{12(4)^{j-1}+3(-1)^{j-1}}{5}} \Big [ 3 + \alpha_3^2(n-1) \Big ]^3 \cr\cr
& = & 2^{\frac{2(4)^{n-m+2}-5-3(-1)^{n-m+1}}{10}} f_3(m)^{\frac{2(4)^{n-m+1}+3(-1)^{n-m+1}}{5}} h_3(m)^{\frac{3(4)^{n-m+1}-3(-1)^{n-m+1}}{5}} \cr\cr 
& & \times \prod _{j=2}^{n-m+1} \Big [ 3 + \alpha_3^2(n+1-j) \Big ]^{\frac{3(4)^{j-1}-3(-1)^{j-1}}{5}} \cr & &
\eeqs
such that Eq. (\ref{f3gen}) is proved by induction. Eq. (\ref{h3gen}) can be established by the same procedure.
\ $\Box$

\bigskip

Let us state the following lemma without proof.

\bigskip

\begin{lemma} \label{lemmachen} If $X(n+1)=\frac {X(n)^2}{c}$ for non-negative integer $n$ and $X_0$ is known, then
\beq
X(n)=\frac {X(0)^{2^n}}{c^{2^n-1}} \ .
\eeq

\end{lemma}

\bigskip

From above lemmas, we have the following bounds for the entropy.

\bigskip

\begin{lemma} \label{lemmasg3b} The entropy for the number of dimer coverings on $SG_3(n)$ is bounded:
\beqs
\frac {-\sqrt{3}\epsilon_3(m)^3}{720(4)^m} & \leq & S_{SG_3} - \Big \{ \frac {2\ln f_3(m)+3\ln h_3(m)+5\ln 2+\ln 3}{10(4)^{m}} + \frac {\sqrt{3}\epsilon_3(m) }{40(4)^m} \Big \} \cr\cr
& \leq & \frac {\sqrt{3}\epsilon_3(m)^2}{ 40(4)^m[2\sqrt{3}-\epsilon_3(m)]} \ ,
\label{Ssg3b}
\eeqs
where $m$ is a positive integer and $\epsilon_3(n)$ is defined as $\alpha_3(n) - \sqrt{3}$. 
\end{lemma}

{\sl Proof} \quad 
As $f_3(n)$ only differs from $s_3(n)$ by a factor of three, which is insignificant in the definition of the entropy, we will substitute $N(G)=f_3(n)$ in Eq. (\ref{sdef}) for $SG_3$ so that 
\beq
S_{SG_3} = \lim_{n\rightarrow\infty} \frac {\ln f_3(n)} {2(4^n+1)} \ .
\label{Ssg3}
\eeq
By Lemma \ref{lemmasg3fh}, we have
\beqs
\ln f_3(n) & = & \frac{2(4)^{n-m}+3(-1)^{n-m}}{5} \ln f_3(m) + \frac{3(4)^{n-m}-3(-1)^{n-m}}{5} \ln h_3(m) \cr\cr
& & + \frac{2(4)^{n-m+1}-5-3(-1)^{n-m}}{10} \ln 2 + \Delta_3(n,m) \ ,
\eeqs
where
\beq
\Delta_3(n,m) = \sum_{j=2}^{n-m} \frac{3(4)^{j-1}-3(-1)^{j-1}}{5} \ln [3+\alpha_3^2(n-j)] \ ,
\label{Delta3}
\eeq
which is bounded as follows.

By Lemma \ref{lemmasg3d}, we know $\epsilon_3(n)$ decreases monotonically to zero for positive integer $n$. $\epsilon_3(1) = \alpha_3(1)-\sqrt{3} = 2-\sqrt{3}$. It is easy to find, by Eq. (\ref{h3overf3}), that 
\beq
\epsilon_3(n+1) = \frac {\epsilon_3(n)^2} {2(\sqrt{3}+\epsilon_3(n))} \ .
\label{epsilon3}
\eeq
Therefore, for any integer $n\geq m$ with $m$ fixed, we have 
\beq
\epsilon_3(n+m)= \frac {\epsilon_3(m)^{2^n}}{(2\sqrt{3})^{2^n-1}} (1+o(n)) 
\label{epsilon3m}
\eeq
by Lemma \ref{lemmachen} and Eq. (\ref{epsilon3}), where $o(n)$ is negative here and $o(n)\rightarrow 0$ as $n\rightarrow \infty$. Now $\Delta_3(n,m)$ in Eq. (\ref{Delta3}) can be rewritten,
\beq
\Delta_3(n,m) = \sum_{j=2}^{n-m} \frac {3(4)^{j-1}-3(-1)^{j-1}}{5} \ln [6+2\sqrt{3}\epsilon_3(n-j)+\epsilon_3(n-j)^2] \ .
\eeq
Since $\epsilon_3(n)$ is small for positive $n$, the logarithmic term can be written as 
\beq
\ln [ 6+2\sqrt{3}\epsilon_3(n-j)+\epsilon_3(n-j)^2] = \ln 6+\frac{\sqrt{3}}{3}\epsilon_3(n-j) \Big [ 1-\frac{\xi_{n,j}\epsilon_3(n-j)^2}{18} \Big ] \ ,
\eeq
where $\xi_{n,j}\in(0,1)$ so that
\beqs
\Delta_3(n,m)
& = & \sum_{j=2}^{n-m}\frac {3(4)^{j-1}-3(-1)^{j-1}} {5} \biggl\{ \ln 6+ \frac {\sqrt{3}\epsilon_3(n-j)}{3} \Big[ 1-\frac {\xi_{n,j}\epsilon_3(n-j)^2}{18} \Big ] \biggr\} \cr\cr
& = & \sum_{j=2}^{n-m}\frac {3\ln 6}{5}[(4)^{j-1}-(-1)^{j-1}] \cr\cr
& & + \sum_{j=2}^{n-m}\frac {\sqrt{3}\epsilon_3(n-j)}{5}[(4)^{j-1}-(-1)^{j-1}] \Big [ 1-\frac {\xi_{n,j}\epsilon_3(n-j)^2}{18} \Big ] \ .
\eeqs
Because the $j=n-m$ term gives the largest contribution for $\Delta_3(n,m)$, it is easy to see that 
\beq
\frac {4^{n-m-1}\sqrt{3}\epsilon_3(m) (1-\frac {\epsilon_3(m)^2}{18})(1+o(n))} {5} \leq \Delta_3(n,m) - \frac {4^{n-m}\ln 6 }{5} (1+o(n)) \ .
\label{Delta3lower}
\eeq
On the other hand, since $\sum_{j=2}^{n-m} \epsilon_3(n-j) 4^{j-1} [1-\xi_{n,j}\epsilon_3(n-j)^2/18]$ is less than $\sum_{i=0}^{n-m-2} 4^{n-m-1} \epsilon_3(m+i)$, we have
\beqs
\Delta_3(n,m) - \frac {4^{n-m}\ln 6 }{5} (1+o(n)) & \leq & \frac {3(4)^{n-m}}{10}(1+o(n)) \sum_{i=0}^{n-m-2} \Big ( \frac {\epsilon_3(m)}{2\sqrt{3}} \Big )^{2^i} \cr\cr
& \leq & \frac {3(4)^{n-m}\epsilon_3(m)(1+o(n))}{10[2\sqrt{3}-\epsilon_3(m)]} \ ,
\label{Delta3upper}
\eeqs
where we use Eq. (\ref{epsilon3m}) and the inequality
\beq
\sum_{j=0}^{n-m-2}x^{2^j} = x+\sum_{j=1}^{n-m-2}x^{2^j} \leq x+\sum_{j=1}^{n-m-2}x^{2j} \leq \frac {x+x^2}{1-x^2} = \frac{x}{1-x}
\label{inequality}
\eeq
for any $0<x<1$. The proof is completed by taking the infinite $n$ limit in Eq. (\ref{Ssg3}).
\ $\Box$

\bigskip

The difference between the upper and lower bounds for $S_{SG_3}$ quickly converges to zero as $m$ increases, and we have the following proposition.

\bigskip

\begin{propo} \label{proposg3} The entropy for the number of dimer coverings on the three-dimensional Sierpinski gasket $SG_3(n)$ in the large $n$ limit is $S_{SG_3}=0.42896389912...$. 

\end{propo}

\bigskip

By Eq. (\ref{epsilon3m}), we know
\beq
\epsilon_3(7) \leq 2\sqrt{3} \Big ( \frac {2-\sqrt{3}}{2\sqrt 3} \Big )^{2^6} \ ,
\eeq
such that $S_{SG_3}$ can be calculated with more than a hundred significant figures accurate when $m$ is equal to seven in Eq. (\ref{Ssg3b}). It is too lengthy to be included here and is available from the authors on request.

\subsection{$SG_4(n)$}

For the four-dimensional Sierpinski gasket $SG_4(n)$, we use the following definitions.

\bigskip

\begin{defi} \label{defisg4} Consider the four-dimensional Sierpinski gasket $SG_4(n)$ at stage $n$. (i) Define $f_4(n)$ as the number of dimer coverings such that the five outmost vertices are vacant. (ii) Define $h_4(n)$ as the number of dimer coverings such that two certain outmost vertices are occupied by dimers and the other three outmost vertices are vacant. (iii) Define $s_4(n)$ as the number of dimer coverings such that one certain outmost vertex is vacant and the other four outmost vertices are occupied by dimers. 
\end{defi}

\bigskip

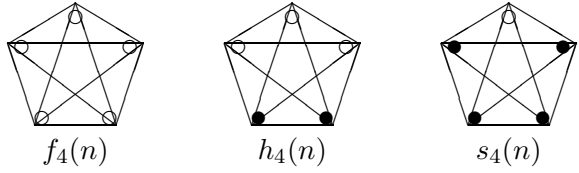
\begin{figure}[htbp]
\unitlength 1.8mm 
\begin{picture}(40,9)
\put(2,0){\line(1,0){6}}
\put(2,0){\line(4,3){8}}
\put(2,0){\line(-1,3){2}}
\put(2,0){\line(1,3){3}}
\put(8,0){\line(-1,3){3}}
\put(8,0){\line(1,3){2}}
\put(8,0){\line(-4,3){8}}
\put(0,6){\line(1,0){10}}
\put(5,9){\line(5,-3){5}}
\put(5,9){\line(-5,-3){5}}
\put(2.5,0.5){\circle{1}}
\put(7.5,0.5){\circle{1}}
\put(1,5.7){\circle{1}}
\put(9,5.7){\circle{1}}
\put(5,8){\circle{1}}
\put(5,-2){\makebox(0,0){$f_4(n)$}}
\put(18,0){\line(1,0){6}}
\put(18,0){\line(4,3){8}}
\put(18,0){\line(-1,3){2}}
\put(18,0){\line(1,3){3}}
\put(24,0){\line(-1,3){3}}
\put(24,0){\line(1,3){2}}
\put(24,0){\line(-4,3){8}}
\put(16,6){\line(1,0){10}}
\put(21,9){\line(5,-3){5}}
\put(21,9){\line(-5,-3){5}}
\put(18.5,0.5){\circle*{1}}
\put(23.5,0.5){\circle*{1}}
\put(17,5.7){\circle{1}}
\put(25,5.7){\circle{1}}
\put(21,8){\circle{1}}
\put(21,-2){\makebox(0,0){$h_4(n)$}}
\put(34,0){\line(1,0){6}}
\put(34,0){\line(4,3){8}}
\put(34,0){\line(-1,3){2}}
\put(34,0){\line(1,3){3}}
\put(40,0){\line(-1,3){3}}
\put(40,0){\line(1,3){2}}
\put(40,0){\line(-4,3){8}}
\put(32,6){\line(1,0){10}}
\put(37,9){\line(5,-3){5}}
\put(37,9){\line(-5,-3){5}}
\put(34.5,0.5){\circle*{1}}
\put(39.5,0.5){\circle*{1}}
\put(33,5.7){\circle*{1}}
\put(41,5.7){\circle*{1}}
\put(37,8){\circle{1}}
\put(37,-2){\makebox(0,0){$s_4(n)$}}
\end{picture}

\vspace*{5mm}
\caption{\footnotesize{Illustration for the dimer coverings $f_4(n)$, $h_4(n)$, $s_4(n)$. Only the five outmost vertices are shown explicitly, where each open circle is vacant and each solid circle is occupied by a dimer.}} 
\label{fhsfign}
\end{figure}

By Eq. (\ref{v}), we have
\beq
v(SG_4(n)) = \frac52 (5^n+1) = 2(5)^n + 3 + \frac12 (5^n-1) 
= 2(5)^n + 3 + \frac12 \sum_{j=1}^n {n \choose j} 4^j \ ,
\eeq
such that the number of vertices for $SG_4(n)$ is always odd. Therefore, we do not have the dimer coverings such that one certain outmost vertices is occupied by a dimer and the other four outmost vertices are vacant, or three certain outmost vertices are occupied by dimers and the other two outmost vertices are vacant, or all five outmost vertices are occupied by dimers.
The quantities $f_4(n)$, $h_4(n)$, and $s_4(n)$ are illustrated in Fig. \ref{fhsfign}, where only the outmost vertices are shown. There are ${5 \choose 2}=10$ equivalent $h_4(n)$ and ${5 \choose 1}=5$ equivalent $s_4(n)$. The initial values at stage zero are again $f_4(0)=1$, $h_4(0)=1$, $s_4(0)=3$. 

\bigskip

\bigskip

We write a computer program to obtain following recursion relations.

\bigskip

\begin{lemma} \label{lemmasg4r} For any non-negative integer $n$,
\beq
f_4(n+1) = 40f_4(n)h_4^3(n)s_4(n) + 24h_4^5(n) \ , 
\label{f4eq}
\eeq
\beq
h_4(n+1) = 24f_4(n)h_4^2(n)s_4^2(n) + 40h_4^4(n)s_4(n) \ , 
\label{h4eq}
\eeq
\beq
s_4(n+1) = 8f_4(n)h_4(n)s_4^3(n) + 56h_4^3(n)s_4^2(n) \ .
\label{s4eq}
\eeq
\end{lemma}

\bigskip

The values of $f_4(n)$, $h_4(n)$, $s_4(n)$ for small $n$ can be evaluated recursively by Eqs. (\ref{f4eq})-(\ref{s4eq}), but they grow exponentially, and do not have simple integer factorizations. To estimate the value of entropy for $SG_4$, we define the ratios
\beq
\alpha_4(n) = \frac{h_4(n)}{f_4(n)} \ , \qquad \beta_4(n) = \frac{s_4(n)}{h_4(n)} \ ,
\label{alpha4def}
\eeq
and their limits
\beq
\alpha_4 \equiv \lim_{n \to \infty} \alpha_4(n) \ , \qquad \beta_4 \equiv \lim_{n \to \infty} \beta_4(n) \ .
\eeq

\bigskip

\begin{lemma} \label{lemmasg4d} Sequence $\{ \alpha_4(n) \}_{n=1}^{\infty}$ decreases monotonically while sequence $\{ \beta_4(n) \}_{n=1}^{\infty}$ increases monotonically. The ratio $\beta_4(n)/\alpha_4(n)$ for positive $n$ increases monotonically to one.
\end{lemma}

{\sl Proof} \quad From Eqs. (\ref{f4eq}) to (\ref{s4eq}), we find
\beq
h_4^2(n+1)-f_4(n+1)s_4(n+1) = 256h_4^4(n)s_4(n) [h_4^2(n)-f_4(n)s_4(n)]^2 \ge 0 \ ,
\eeq
such that $\alpha_4(n) \ge \beta_4(n)$ for $n>0$. It follows that for positive $n$,
\beq
\frac{h_4(n)}{f_4(n)} - \frac{h_4(n+1)}{f_4(n+1)} = \frac{24h_4^2(n)[h_4^4(n)-f_4^2(n)s_4^2(n)]}{f_4(n)f_4(n+1)} \ge 0 \ ,
\eeq
which shows that $\alpha_4(n)$ decreases, and 
\beq
\frac{s_4(n)}{h_4(n)} - \frac{s_4(n+1)}{h_4(n+1)} = \frac{16h_4^2(n)s_4^2(n)[f_4(n)s_4(n)-h_4^2(n)]}{h_4(n)h_4(n+1)} \le 0 \ ,
\eeq
which shows that $\beta_4(n)$ increases. From Eqs. (\ref{f4eq}) to (\ref{s4eq}), we have
\beq
\frac{h_4(n+1)}{f_4(n+1)} = \frac{\beta_4(n)[3\beta_4(n)+5\alpha_4(n)]}{5\beta_4(n)+3\alpha_4(n)} \ ,
\label{h4overf4}
\eeq
and
\beq
\frac{s_4(n+1)}{h_4(n+1)} = \frac{\beta_4(n)[\beta_4(n)+7\alpha_4(n)]}{3\beta_4(n)+5\alpha_4(n)} \ ,
\label{s4overh4}
\eeq
which leads to $\alpha_4=\beta_4$ by taking the limit $n \to \infty$. The actual value of $\alpha_4$ and $\beta_4$ cannot be obtained by solving these equations. The numerical results give
\beq
\alpha_4 = \beta_4 = 0.850772150002...
\label{alpha4value}
\eeq
where more than a hundred significant figures can be evaluated when stage $n$ in Eq. (\ref{alpha4def}) is equal to seven.
\ $\Box$

\bigskip

The general expressions for $h_4(n)$ and $s_4(n)$ in terms of quantities at stage $m<n$ can be written as follows.

\bigskip

\begin{lemma} \label{lemmasg4hs} For a non-negative integer $m$ and any positive integer $n > m$, 
\beqs
h_4(n) & = & 2^{\frac{3(5)^{n-m}-3}{4}} h_4(m)^{\frac{3(5)^{n-m}+1}{4}} s_4(m)^{\frac{5^{n-m}-1}{4}} \cr\cr & & 
\times \prod _{i=1}^{n-m} \Big [ 5 + 3\frac{\beta_4(n-i)}{\alpha_4(n-i)} \Big ]^{\frac{3(5)^{i-1}+1}{4}} \prod _{j=2}^{n-m} \Big [ 7 + \frac{\beta_4(n-j)}{\alpha_4(n-j)} \Big ]^{\frac{5^{j-1}-1}{4}} \ , \cr & &
\label{h4gen}
\eeqs
\beqs
s_4(n) & = & 2^{\frac{3(5)^{n-m}-3}{4}} h_4(m)^{\frac{3(5)^{n-m}-3}{4}} s_4(m)^{\frac{5^{n-m}+3}{4}} \cr\cr & & 
\times \prod _{i=2}^{n-m} \Big [ 5 + 3\frac{\beta_4(n-i)}{\alpha_4(n-i)} \Big ]^{\frac{3(5)^{i-1}-3}{4}} \prod _{j=1}^{n-m} \Big [ 7 + \frac{\beta_4(n-j)}{\alpha_4(n-j)} \Big ]^{\frac{5^{j-1}+3}{4}} \ . \cr & &
\label{s4gen}
\eeqs
Here when $n-m=1$, the products with lower limit two are defined to be one.
\end{lemma}

{\sl Proof} \quad 
It is clear that Eqs. (\ref{h4gen}) and (\ref{s4gen}) are valid for $n=m+1$ since $h_4(m+1)=8h_4^4(m)s_4(m)[5+3\beta_4(m)/\alpha_4(m)]$ and $s_4(m+1)=8h_4^3(m)s_4^2(m)[7+\beta_4(m)/\alpha_4(m)]$ by Eqs. (\ref{h4eq}) and (\ref{s4eq}), respectively. Consider Eq. (\ref{h4gen}) holds for a certain positive integer $n>m$, then 
\beqs
\lefteqn{h_4(n+1) = 8 h_4^4(n) s_4(n) \Big [ 5 + 3\frac{\beta_4(n)}{\alpha_4(n)} \Big ]} \cr\cr
& = & 8 \Big [ 5 + 3\frac{\beta_4(n)}{\alpha_4(n)} \Big ] \times 2^{3(5)^{n-m}-3} h_4(m)^{3(5)^{n-m}+1} s_4(m)^{5^{n-m}-1} \prod _{i=1}^{n-m} \Big [ 5 + 3\frac{\beta_4(n-i)}{\alpha_4(n-i)} \Big ]^{3(5)^{i-1}+1} \cr\cr & & 
\times \prod _{j=2}^{n-m} \Big [ 7 + \frac{\beta_4(n-j)}{\alpha_4(n-j)} \Big ]^{5^{j-1}-1}  2^{\frac{3(5)^{n-m}-3}{4}} h_4(m)^{\frac{3(5)^{n-m}-3}{4}} s_4(m)^{\frac{5^{n-m}+3}{4}} \cr\cr & & 
\times \prod _{i=2}^{n-m} \Big [ 5 + 3\frac{\beta_4(n-i)}{\alpha_4(n-i)} \Big ]^{\frac{3(5)^{i-1}-3}{4}} \prod _{j=1}^{n-m} \Big [ 7 + \frac{\beta_4(n-j)}{\alpha_4(n-j)} \Big ]^{\frac{5^{j-1}+3}{4}}  \cr\cr
& = & 2^{\frac{3(5)^{n-m+1}-3}{4}} h_4(m)^{\frac{3(5)^{n-m+1}+1}{4}} s_4(m)^{\frac{5^{n-m+1}-1}{4}} \prod _{i=2}^{n-m} \Big [ 5 + 3\frac{\beta_4(n-i)}{\alpha_4(n-i)} \Big ]^{\frac{3(5)^i+1}{4}} \cr\cr & & 
\times \Big [ 5 + 3\frac{\beta_4(n-1)}{\alpha_4(n-1)} \Big ]^4 \Big [ 5 + 3\frac{\beta_4(n)}{\alpha_4(n)} \Big ]  \prod _{j=2}^{n-m} \Big [ 7 + \frac{\beta_4(n-j)}{\alpha_4(n-j)} \Big ]^{\frac{5^j-1}{4}} \Big [ 7 + \frac{\beta_4(n-1)}{\alpha_4(n-1)} \Big ] \cr\cr
& = & 2^{\frac{3(5)^{n-m+1}-3}{4}} h_4(m)^{\frac{3(5)^{n-m+1}+1}{4}} s_4(m)^{\frac{5^{n-m+1}-1}{4}} \prod _{i=1}^{n-m+1} \Big [ 5 + 3\frac{\beta_4(n+1-i)}{\alpha_4(n+1-i)} \Big ]^{\frac{3(5)^{i-1}+1}{4}} \cr\cr & & 
\times  \prod _{j=2}^{n-m+1} \Big [ 7 + \frac{\beta_4(n+1-j)}{\alpha_4(n+1-j)} \Big ]^{\frac{5^{j-1}-1}{4}} \ ,
\eeqs
such that Eq. (\ref{h4gen}) is proved by induction. Eq. (\ref{s4gen}) can be established by the same procedure.
\ $\Box$

\bigskip

From above lemmas, we have the following bounds for the entropy.

\bigskip

\begin{lemma} \label{lemmasg4b} The entropy for the number of dimer coverings on $SG_4(n)$ is bounded:
\beq
-\frac {7\epsilon_4(m)^2}{640(5)^m[1-\frac {\epsilon_4(m)}{16}]} \leq S_{SG_4} - \Big \{ \frac {3\ln h_4(m)+\ln s_4(m)+6\ln 2}{10(5)^{m}} - \frac {\epsilon_4(m)}{40(5)^{m}} \Big \} \leq 0 \ ,
\label{Ssg4b}
\eeq
where $m$ is a positive integer and $\epsilon_4(n)$ is defined as $1-\beta_4(n)/\alpha_4(n)$. 
\end{lemma}

{\sl Proof} \quad 
We substitute $N(G)=s_4(n)$ in Eq. (\ref{sdef}) for $SG_4$ so that 
\beq
S_{SG_4} = \lim_{n\rightarrow\infty} \frac {\ln s_4(n)} {5(5^n+1)/2} \ .
\label{Ssg4}
\eeq 
By Lemma \ref{lemmasg4hs}, we have
\beqs
\ln s_4(n) & = & \frac{3(5)^{n-m}-3}{4} \ln h_4(m) + \frac{5^{n-m}+3}{4} \ln s_4(m) + \frac{3(5)^{n-m}-3}{4} \ln 2 + \Delta_4(n,m) \ , \cr & &
\eeqs
where
\beqs
\Delta_4(n,m) & = & \sum_{i=2}^{n-m} \frac{3(5)^{i-1}-3}{4} \ln \Big [ 5 + 3\frac{\beta_4(n-i)}{\alpha_4(n-i)} \Big ] + \sum_{j=1}^{n-m} \frac{5^{j-1}+3}{4} \ln \Big [ 7 + \frac{\beta_4(n-j)}{\alpha_4(n-j)} \Big ] \ , \cr & &
\label{Delta4}
\eeqs
which is bounded as follows.

By Lemma \ref{lemmasg4d}, we know $\epsilon_4(n)$ decreases monotonically to zero for positive integer $n$. $\epsilon_4(1) = 1 - \beta_4(1)/\alpha_4(1) = 4/49$. It is easy to find, by Eqs. (\ref{h4overf4}) and (\ref{s4overh4}), that 
\beq
\epsilon_4(n+1) = \Big [ \frac {2\epsilon_4(n)}{8-3\epsilon_4(n)} \Big ]^2 \ .
\label{epsilon4}
\eeq
Therefore, for any integer $n\geq m$ with $m$ fixed, we have 
\beq
\epsilon_4(n+m) = \frac {\epsilon_4(m)^{2^n}}{16^{2^n-1}} (1+o(n)) \ .
\label{epsilon4m}
\eeq
Now $\Delta_4(n,m)$ in Eq. (\ref{Delta4}) can be rewritten,
\beqs
\Delta_4(n,m) & = &\sum_{i=2}^{n-m} \frac {3[ (5)^{i-1}-1]}{4} \Big [ \ln 8+\ln (1-\frac {3\epsilon_4(n-i)}{8} ) \Big ] \cr\cr
& & +\sum_{j=1}^{n-m}\frac {(5)^{j-1}+3}{4} \Big [ \ln 8+\ln (1-\frac {\epsilon_4(n-j)}{8} ) \Big ] \cr\cr
& = & \sum_{i=2}^{n-m} \frac {3[ (5)^{i-1}-1]}{4} \Big \{\ln 8-\frac {3\epsilon_4(n-i)}{8} \Big [1+\frac {3\xi_{n,j}\epsilon_4(n-j)}8 \Big ] \Big \} \cr\cr
& & +\sum_{j=1}^{n-m}\frac {(5)^{j-1}+3}{4} \Big \{\ln 8-\frac {\epsilon_4(n-j)}{8} \Big [1+\frac {\xi_{n,j}'\epsilon_4(n-j)}8 \Big ] \Big \} \ , \cr & &
\eeqs
where $\xi_{n,j},\xi_{n,j}'\in(0,1)$. It is easy to see that
\beq
\Delta_4(n,m) - \frac {(3\ln 2)5^{n-m}}{4} (1+o(n)) \leq -\frac {5^{n-m}\epsilon_4(m)}{16}(1+o(n)) \ .
\label{Delta4upper}
\eeq
By Eqs. (\ref{inequality}) and (\ref{epsilon4m}), we have
\beqs
& & \Delta_4(n,m) - \frac {(3\ln 2)5^{n-m}}{4} (1+o(n)) \cr\cr
& \geq & -5^{n-m}(1+o(n)) \Big ( 1+\frac{3\epsilon_4(m)}8 \Big ) \sum_{j=0}^{n-m-1} \Big ( \frac {\epsilon_4(m)}{16} \Big )^{2^j} \cr\cr
& \geq & -\frac {\epsilon_4(m)5^{n-m}(1+o(n)) \Big ( 1+\frac {3\epsilon_4(m)}8 \Big ) }{16-\epsilon_4(m)} \ .
\label{Delta4lower}
\eeqs
The proof is completed by taking the infinite $n$ limit in Eq. (\ref{Ssg4}).
\ $\Box$

\bigskip

The difference between the upper and lower bounds for $S_{SG_4}$ quickly converges to zero as $m$ increases, and we have the following proposition.

\bigskip

\begin{propo} \label{proposg4} The entropy for the number of dimer coverings on the four-dimensional Sierpinski gasket $SG_4(n)$ in the large $n$ limit is $S_{SG_4}=0.56337479920...$. 

\end{propo}

\bigskip

The numerical value of $S_{SG_4}$ can be calculated with more than a hundred significant figures accurate when $m$ in Eq. (\ref{Ssg4b}) is equal to six. It is too lengthy to be included here and is available from the authors on request.

\subsection{$SG_5(n)$}

For the five-dimensional Sierpinski gasket $SG_5(n)$, we use the following definitions.

\bigskip

\begin{defi} \label{defisg5} Consider the five-dimensional Sierpinski gasket $SG_5(n)$ at stage $n$. (i) Define $f_5(n)$ as the number of dimer coverings such that the six outmost vertices are vacant. (ii) Define $g_5(n)$ as the number of dimer coverings such that one certain outmost vertex is occupied by a dimer and the other five outmost vertices are vacant. (iii) Define $h_5(n)$ as the number of dimer coverings such that two certain outmost vertices are occupied by dimers and the other four outmost vertices are vacant. (iv) Define $r_5(n)$ as the number of dimer coverings such that three certain outmost vertices are occupied by dimers and the other three outmost vertices are vacant. (v) Define $s_5(n)$ as the number of dimer coverings such that two certain outmost vertices are vacant and the other four outmost vertices are occupied by dimers. (vi) Define $t_5(n)$ as the number of dimer coverings such that one certain outmost vertex is vacant and the other five outmost vertices are occupied by dimers. (vii) Define $u_5(n)$ as the number of dimer coverings such that all six outmost vertices are occupied by dimers. 
\end{defi}

\bigskip

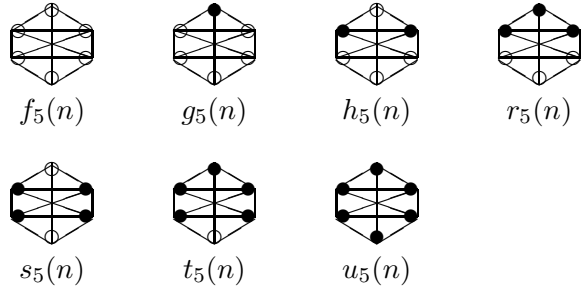
\begin{figure}[htbp]
\unitlength 1.8mm 
\begin{picture}(42,6)
\put(3,0){\line(5,3){3}}
\put(3,0){\line(-5,3){3}}
\put(3,0){\line(0,1){6}}
\put(0,2){\line(1,0){6}}
\put(0,2){\line(3,1){6}}
\put(0,2){\line(0,1){2}}
\put(0,4){\line(3,-1){6}}
\put(0,4){\line(1,0){6}}
\put(0,4){\line(5,3){3}}
\put(6,2){\line(0,1){2}}
\put(6,4){\line(-5,3){3}}
\put(3,0.5){\circle{1}}
\put(0.5,2){\circle{1}}
\put(0.5,4){\circle{1}}
\put(3,5.5){\circle{1}}
\put(5.5,4){\circle{1}}
\put(5.5,2){\circle{1}}
\put(3,-2){\makebox(0,0){$f_5(n)$}}
\put(15,0){\line(5,3){3}}
\put(15,0){\line(-5,3){3}}
\put(15,0){\line(0,1){6}}
\put(12,2){\line(1,0){6}}
\put(12,2){\line(3,1){6}}
\put(12,2){\line(0,1){2}}
\put(12,4){\line(3,-1){6}}
\put(12,4){\line(1,0){6}}
\put(12,4){\line(5,3){3}}
\put(18,2){\line(0,1){2}}
\put(18,4){\line(-5,3){3}}
\put(15,0.5){\circle{1}}
\put(12.5,2){\circle{1}}
\put(12.5,4){\circle{1}}
\put(15,5.5){\circle*{1}}
\put(17.5,4){\circle{1}}
\put(17.5,2){\circle{1}}
\put(15,-2){\makebox(0,0){$g_5(n)$}}
\put(27,0){\line(5,3){3}}
\put(27,0){\line(-5,3){3}}
\put(27,0){\line(0,1){6}}
\put(24,2){\line(1,0){6}}
\put(24,2){\line(3,1){6}}
\put(24,2){\line(0,1){2}}
\put(24,4){\line(3,-1){6}}
\put(24,4){\line(1,0){6}}
\put(24,4){\line(5,3){3}}
\put(30,2){\line(0,1){2}}
\put(30,4){\line(-5,3){3}}
\put(27,0.5){\circle{1}}
\put(24.5,2){\circle{1}}
\put(24.5,4){\circle*{1}}
\put(27,5.5){\circle{1}}
\put(29.5,4){\circle*{1}}
\put(29.5,2){\circle{1}}
\put(27,-2){\makebox(0,0){$h_5(n)$}}
\put(39,0){\line(5,3){3}}
\put(39,0){\line(-5,3){3}}
\put(39,0){\line(0,1){6}}
\put(36,2){\line(1,0){6}}
\put(36,2){\line(3,1){6}}
\put(36,2){\line(0,1){2}}
\put(36,4){\line(3,-1){6}}
\put(36,4){\line(1,0){6}}
\put(36,4){\line(5,3){3}}
\put(42,2){\line(0,1){2}}
\put(42,4){\line(-5,3){3}}
\put(39,0.5){\circle{1}}
\put(36.5,2){\circle{1}}
\put(36.5,4){\circle*{1}}
\put(39,5.5){\circle*{1}}
\put(41.5,4){\circle*{1}}
\put(41.5,2){\circle{1}}
\put(39,-2){\makebox(0,0){$r_5(n)$}}
\end{picture}

\vspace*{10mm}

\begin{picture}(42,6)
\put(3,0){\line(5,3){3}}
\put(3,0){\line(-5,3){3}}
\put(3,0){\line(0,1){6}}
\put(0,2){\line(1,0){6}}
\put(0,2){\line(3,1){6}}
\put(0,2){\line(0,1){2}}
\put(0,4){\line(3,-1){6}}
\put(0,4){\line(1,0){6}}
\put(0,4){\line(5,3){3}}
\put(6,2){\line(0,1){2}}
\put(6,4){\line(-5,3){3}}
\put(3,0.5){\circle{1}}
\put(0.5,2){\circle*{1}}
\put(0.5,4){\circle*{1}}
\put(3,5.5){\circle{1}}
\put(5.5,4){\circle*{1}}
\put(5.5,2){\circle*{1}}
\put(3,-2){\makebox(0,0){$s_5(n)$}}
\put(15,0){\line(5,3){3}}
\put(15,0){\line(-5,3){3}}
\put(15,0){\line(0,1){6}}
\put(12,2){\line(1,0){6}}
\put(12,2){\line(3,1){6}}
\put(12,2){\line(0,1){2}}
\put(12,4){\line(3,-1){6}}
\put(12,4){\line(1,0){6}}
\put(12,4){\line(5,3){3}}
\put(18,2){\line(0,1){2}}
\put(18,4){\line(-5,3){3}}
\put(15,0.5){\circle{1}}
\put(12.5,2){\circle*{1}}
\put(12.5,4){\circle*{1}}
\put(15,5.5){\circle*{1}}
\put(17.5,4){\circle*{1}}
\put(17.5,2){\circle*{1}}
\put(15,-2){\makebox(0,0){$t_5(n)$}}
\put(27,0){\line(5,3){3}}
\put(27,0){\line(-5,3){3}}
\put(27,0){\line(0,1){6}}
\put(24,2){\line(1,0){6}}
\put(24,2){\line(3,1){6}}
\put(24,2){\line(0,1){2}}
\put(24,4){\line(3,-1){6}}
\put(24,4){\line(1,0){6}}
\put(24,4){\line(5,3){3}}
\put(30,2){\line(0,1){2}}
\put(30,4){\line(-5,3){3}}
\put(27,0.5){\circle*{1}}
\put(24.5,2){\circle*{1}}
\put(24.5,4){\circle*{1}}
\put(27,5.5){\circle*{1}}
\put(29.5,4){\circle*{1}}
\put(29.5,2){\circle*{1}}
\put(27,-2){\makebox(0,0){$u_5(n)$}}
\end{picture}

\vspace*{5mm}
\caption{\footnotesize{Illustration for the dimer coverings $f_5(n)$, $g_5(n)$, $h_5(n)$, $r_5(n)$, $s_5(n)$, $t_5(n)$, $u_5(n)$. Only the six outmost vertices are shown explicitly, where each open circle is vacant and each solid circle is occupied by a dimer.}} 
\label{fghrstufig}
\end{figure}

The quantities $f_5(n)$, $g_5(n)$, $h_5(n)$, $r_5(n)$, $s_5(n)$, $t_5(n)$ and $u_5(n)$ are illustrated in Fig. \ref{fghrstufig}, where only the outmost vertices are shown.
The initial values are $f_5(0)=1$, $g_5(0)=0$, $h_5(0)=1$, $r_5(0)=0$, $s_5(0)=3$, $t_5(0)=0$, $u_5(0)=15$. For the five-dimensional Sierpinski gasket $SG_5(n)$, the number of vertices is equal to six for $n=0$ and odd for all positive integer $n$ by Eq. (\ref{v}). Therefore, $f_5(n)$, $h_5(n)$, $s_5(n)$, $u_5(n)$ are always zero for positive integer $n$. There are ${6 \choose 1}=6$ equivalent $g_5(n)$ and $t_5(n)$, and ${6 \choose 3}=20$ equivalent $r_5(n)$. 

\bigskip

\bigskip

We write a computer program to obtain following recursion relations.

\bigskip

\begin{lemma} \label{lemmasg5r} For any positive integer $n$,
\beq
g_5(n+1) = 40g_5^3(n)r_5(n)t_5^2(n) + 560g_5^2(n)r_5^3(n)t_5(n) + 424g_5(n)r_5^5(n) \ , 
\label{g5eq}
\eeq
\beq
r_5(n+1) = 4g_5^3(n)t_5^3(n) + 252g_5^2(n)r_5^2(n)t_5^2(n) + 636g_5(n)r_5^4(n)t_5(n) + 132r_5^6(n) \ , 
\label{r5eq}
\eeq
\beq
t_5(n+1) = 40g_5^2(n)r_5(n)t_5^3(n) + 560g_5(n)r_5^3(n)t_5^2(n) + 424r_5^5(n)t_5(n) \ ,
\label{t5eq}
\eeq
and for $n=0$,
\beqs
g_5(1) & = & 280f_5(0)h_5^2(0)s_5^3(0) + 40f_5(0)h_5^3(0)s_5(0)u_5(0) + 680h_5^4(0)s_5^2(0) + 24h_5^5(0)u_5(0) \cr\cr
& = & 15840 \ , 
\label{g51eq}
\eeqs
\beqs
r_5(1) & = & 72f_5(0)h_5^2(0)s_5^2(0)u_5(0) + 120f_5(0)h_5(0)s_5^4(0) + 712h_5^3(0)s_5^3(0) + 120h_5^4(0)s_5(0)u_5(0) \cr\cr
& = & 44064 \ , 
\label{r51eq}
\eeqs
\beqs
t_5(1) & = & 40f_5(0)h_5(0)s_5^3(0)u_5(0) + 280h_5^3(0)s_5^2(0)u_5(0) + 24f_5(0)s_5^5(0) + 680h_5^2(0)s_5^4(0) \cr\cr
& = & 114912 \ . 
\label{t51eq}
\eeqs
\end{lemma}

\bigskip

The values of $g_4(n)$, $r_4(n)$, $t_4(n)$ for small positive $n$ can be evaluated recursively by Eqs. (\ref{g5eq})-(\ref{t5eq}), but they grow exponentially, and do not have simple integer factorizations. To estimate the value of entropy for $SG_5$, we define the ratio
\beq
\alpha_5(n) = \frac{r_5(n)}{g_5(n)} \ ,
\label{alpha5def}
\eeq
and its limit
\beq
\alpha_5 \equiv \lim_{n \to \infty} \alpha_5(n) \ .
\eeq

\bigskip

\begin{lemma} \label{lemmasg5d} Sequence $\{ \alpha_5(n) \}_{n=1}^{\infty}$ decreases monotonically. The limit $\alpha_5$ is equal to $\sqrt{399/55}$.
\end{lemma}

{\sl Proof} \quad From Eqs. (\ref{g5eq}) and (\ref{t5eq}), the ratio $t_5(n)/g_5(n)$ is invariant. Defined the ratio as $c$, then
\beq
c = \frac{t_5(1)}{g_5(1)} = \frac{399}{55} \ .
\eeq
Eqs. (\ref{g5eq}) and (\ref{r5eq}) can be modified to be
\beq
g_5(n+1) = 8 g_5^5(n) r_5(n) P_5(n) \ ,
\label{g5eqn}
\eeq
\beq
r_5(n+1) = 4 g_5^6(n) Q_5(n) \ ,
\label{r5eqn}
\eeq
where 
\beq
P_5(n) = 5c^2 + 70c\alpha_5^2(n) + 53\alpha_5^4(n) \ ,
\label{P}
\eeq
\beq
Q_5(n) = c^3 + 63c^2\alpha_5^2(n) + 159c\alpha_5^4(n) + 33\alpha_5^6(n) \ .
\label{Q}
\eeq
It is clear that $\alpha_5(n)$ is bounded below by $\sqrt{c}$ because
\beqs
& & r_5^2(n+1)-g_5(n+1)t_5(n+1) = r_5^2(n+1)-cg_5^2(n+1) \cr\cr
& = & 16[r_5^2(n)-cg_5^2(n)] [c^4g_5^8(n) + 28c^3g_5^6(n)r_5^2(n) + 1542c^2g_5^4(n)r_5^4(n) \cr\cr
& & + 1436cg_5^2(n)r_5^6(n) + 1089r_5^8(n) ] \ge 0 
\eeqs
by induction. It follows that $\alpha_5(n)$ decreases for positive $n$ because
\beqs
& & \frac{r_5(n)}{g_5(n)} - \frac{r_5(n+1)}{g_5(n+1)}  \cr\cr
& = & \frac{4}{g_5(n+1)} \Big \{19r_5^4(n)[r_5^2(n)-cg_5^2(n)] + 53r_5^2(n)[r_5^4(n)-c^2g_5^4(n)] + r_5^6(n)-c^3g_5^6(n) \Big \} \ge 0 \ , \cr & & 
\eeqs
which implies that the limit $\alpha_5$ exists. From Eqs. (\ref{g5eqn}) and (\ref{r5eqn}), we have
\beq
\frac{r_5(n+1)}{g_5(n+1)} = \frac{c^3+63c^2\alpha_5^2(n)+159c\alpha_5^4(n)+33\alpha_5^6(n)} {2\alpha_5(n)[5c^2+70c\alpha_5^2(n)+53\alpha_5^4(n)]} \ .
\label{r5overg5}
\eeq
By taking the large $n$ limit in Eq. (\ref{r5overg5}) and the requirement that $\alpha_5$ must be real and positive, $\alpha_5$ is solved to be $\sqrt{c}$ .
\ $\Box$

\bigskip

The general expressions for $g_5(n)$ and $r_5(n)$ in terms of quantities at stage $m<n$ can be written as follows.

\bigskip

\begin{lemma} \label{lemmasg5gr} For a non-negative integer $m$ and any positive integer $n > m$, 
\beqs
g_5(n) & = & 2^{\frac{8(6)^{n-m}-7-(-1)^{n-m}}{14}} g_5(m)^{\frac{6^{n-m+1}+(-1)^{n-m}}{7}} r_5(m)^{\frac{6^{n-m}-(-1)^{n-m}}{7}} \cr\cr & & 
\times \prod _{i=1}^{n-m} P_5(n-i)^{\frac{6^i-(-1)^i}{7}} \prod _{j=2}^{n-m} Q_5(n-j)^{\frac{6^{j-1}+(-1)^j}{7}} \ ,
\label{g5gen}
\eeqs
\beqs
r_5(n) & = & 2^{\frac{4(6)^{n-m}-7+3(-1)^{n-m}}{7}} g_5(m)^{\frac{6^{n-m+1}-6(-1)^{n-m}}{7}} r_5(m)^{\frac{6^{n-m}+6(-1)^{n-m}}{7}} \cr\cr & & 
\times \prod _{i=2}^{n-m} P_5(n-i)^{\frac{6^i+6(-1)^i}{7}} \prod _{j=1}^{n-m} Q_5(n-j)^{\frac{6^{j-1}-6(-1)^j}{7}} \ .
\label{r5gen}
\eeqs
Here when $n-m=1$, the products with lower limit two are defined to be one.
\end{lemma}

{\sl Proof} \quad 
It is clear that Eqs. (\ref{g5gen}) and (\ref{r5gen}) are valid for $n=m+1$ since $g_5(m+1)=8g_5^5(m)r_5(m)P_5(m)$ and $r_5(m+1)=4g_5^6(m)Q_5(m)$ by Eqs. (\ref{g5eqn}) and (\ref{r5eqn}), respectively. Consider Eq. (\ref{g5gen}) holds for a certain positive integer $n>m$, then 
\beqs
\lefteqn{g_5(n+1) = 8 g_5^5(n) r_5(n) P_5(n)} \cr\cr
& = & 8 \times 2^{\frac{40(6)^{n-m}-35-5(-1)^{n-m}}{14}} g_5(m)^{\frac{5(6)^{n-m+1}+5(-1)^{n-m}}{7}} r_5(m)^{\frac{5(6)^{n-m}-5(-1)^{n-m}}{7}} \cr\cr & & 
\times \prod _{i=1}^{n-m} P_5(n-i)^{\frac{5(6)^i-5(-1)^i}{7}} \prod _{j=2}^{n-m} Q_5(n-j)^{\frac{5(6)^{j-1}+5(-1)^j}{7}} \cr\cr
& & \times 2^{\frac{4(6)^{n-m}-7+3(-1)^{n-m}}{7}} g_5(m)^{\frac{6^{n-m+1}-6(-1)^{n-m}}{7}} r_5(m)^{\frac{6^{n-m}+6(-1)^{n-m}}{7}} \cr\cr & & 
\times \prod _{i=2}^{n-m} P_5(n-i)^{\frac{6^i+6(-1)^i}{7}} \prod _{j=1}^{n-m} Q_5(n-j)^{\frac{6^{j-1}-6(-1)^j}{7}} P_5(n)  \cr\cr
& = & 2^{\frac{8(6)^{n-m+1}-7+(-1)^{n-m}}{14}} g_5(m)^{\frac{6^{n-m+2}-(-1)^{n-m}}{7}} r_5(m)^{\frac{6^{n-m+1}+(-1)^{n-m}}{7}} \cr\cr & & 
\times \prod _{i=2}^{n-m} P_5(n-i)^{\frac{6^{i+1}+(-1)^i}{7}} P_5(n-1)^5 P_5(n) \prod _{j=2}^{n-m} Q_5(n-j)^{\frac{6^j-(-1)^j}{7}} Q_5(n-1) \cr\cr
& = & 2^{\frac{8(6)^{n-m+1}-7-(-1)^{n-m+1}}{14}} g_5(m)^{\frac{6^{n-m+2}+(-1)^{n-m+1}}{7}} r_5(m)^{\frac{6^{n-m+1}-(-1)^{n-m+1}}{7}} \cr\cr & & 
\times \prod _{i=1}^{n-m+1} P_5(n+1-i)^{\frac{6^i-(-1)^i}{7}} \prod _{j=2}^{n-m+1} Q_5(n+1-j)^{\frac{6^{j-1}+(-1)^j}{7}} \ ,
\eeqs
such that Eq. (\ref{g5gen}) is proved by induction. Eq. (\ref{r5gen}) can be established by the same procedure.
\ $\Box$

\bigskip

From above lemmas, we have the following bounds for the entropy.

\bigskip

\begin{lemma} \label{lemmasg5b} The entropy for the number of dimer coverings on $SG_5(n)$ is bounded:
\beqs
0 & \leq & S_{SG_5} - \Big \{ \frac {2\ln g_5(m)}{7(6)^m}+\frac {\ln r_5(m)}{21(6)^m}+\frac {14\ln 2}{21(6)^m}+\frac {\ln c}{7(6)^m}+\frac {9\epsilon_5(m)}{56\sqrt{c}(6)^{m}} \Big \} \cr\cr
& \leq & \frac { 279\epsilon_5(m)^2 }{448c(6)^m \Big [1-\frac {\epsilon_5(m)}{8\sqrt{c}} \Big ]} \ ,
\label{Ssg5b}
\eeqs
where $m$ is a positive integer and $\epsilon_5(n)$ is defined as $\alpha_5(n) - \sqrt{c}$. 
\end{lemma}

{\sl Proof} \quad 
As $u_5(n)$ is exactly zero for all positive $n$, we will substitute $N(G)=g_5(n)$ in Eq. (\ref{sdef}) for $SG_5$ so that 
\beq
S_{SG_5} = \lim_{n\rightarrow\infty} \frac {\ln g_5(n)} {3(6^n+1)} \ .
\label{Ssg5}
\eeq
By Lemma \ref{lemmasg5gr}, we have
\beqs
\ln g_5(n) & = & \frac{6^{n-m+1}+(-1)^{n-m}}{7} \ln g_5(m) + \frac{6^{n-m}-(-1)^{n-m}}{7} \ln r_5(m) \cr\cr
& & + \frac{8(6)^{n-m}-7-(-1)^{n-m}}{14} \ln 2 + \Delta_5(n,m) \ ,
\eeqs
where
\beq
\Delta_5(n,m) = \sum_{i=1}^{n-m} \frac{6^i-(-1)^i}{7} \ln P_5(n-i) + \sum_{j=2}^{n-m} \frac{6^{j-1}+(-1)^j}{7} \ln Q_5(n-j) \ ,
\label{Delta5}
\eeq
which is bounded as follows.

By Lemma \ref{lemmasg5d}, we know $\epsilon_5(n)$ decreases monotonically to zero for positive integer $n$. $\epsilon_5(1) = \alpha_5(1)-\sqrt{c} = 153/55-\sqrt{399/55}$. It is easy to find, by Eq. (\ref{r5overg5}), that 
\beq
\epsilon_5(n+1) = \frac {\epsilon_5(n)^2} {8\sqrt{c}} (1+o(n)) \ .
\label{epsilon5}
\eeq
Therefore, for any integer $n\geq m$ with $m$ fixed, we have 
\beq
\epsilon_5(m+n) = \frac {\epsilon_5(m)^{2^n}} {(8\sqrt{c})^{2^n-1}} (1+o(n)) \ .   
\label{epsilon5m}
\eeq
From Eqs. (\ref{P}) and (\ref{Q}), we have
\beq
P_5(n+m) = 128c^2+352c\sqrt{c}\epsilon_5(n+m) (1+o(n)) \ ,
\eeq
and
\beq
Q_5(n+m) = 256c^3+960c^2\sqrt{c}\epsilon_5(n+m) (1+o(n)) \ ,
\eeq
such that
\beq
\ln P_5(n+m) = 7\ln 2+2 \ln c+ (1+o(n)) \ln \Big( 1+\frac {11\epsilon_5(n+m)}{4\sqrt{c}} \Big ) \ ,
\eeq
\beq
\ln Q_5(n+m) = 8\ln 2+3 \ln c+ (1+o(n)) \ln \Big( 1+\frac {15\epsilon_5(n+m)}{4\sqrt{c}} \Big ) \ .
\eeq
Now $\Delta_5(n,m)$ in Eq. (\ref{Delta5}) can be rewritten,
\beqs
\Delta_5(n,m) & = & \sum_{i=1}^{n-m}\frac {(6)^{i}-(-1)^i}{7}[7\ln 2+2\ln c] \cr\cr
& & +\sum_{i=1}^{n-m}\frac {(6)^{i}-(-1)^{i}}{7} (1+o(n)) \ln 
\Big( 1+\frac {11\epsilon_5(n-i)}{4\sqrt{c}} \Big) \cr\cr
& & +\sum_{j=2}^{n-m}\frac {(6)^{j-1}+(-1)^j}{7}[8\ln 2+3\ln c] \cr\cr
& & +\sum_{j=2}^{n-m}\frac {(6)^{j-1}-(-1)^{j}}{7} (1+o(n)) \ln 
\Big(1+\frac {15\epsilon_5(n-j) }{ 4\sqrt{c} } \Big) \ . \cr & &
\eeqs
It is easy to see that
\beq
\Delta_5(n,m) - \Big [ \frac {10\ln 2}{7}+\frac {3\ln c}{7} \Big ] 6^{n-m} (1+o(n)) \geq \frac {27 \epsilon_5(m)}{56\sqrt{c}}6^{n-m}(1-o(n)) \ .
\label{Delta5lower}
\eeq
By Eqs. (\ref{inequality}) and (\ref{epsilon5m}), we have
\beqs
& & \Delta_5(n,m) - \Big [ \frac {10\ln 2}{7}+\frac {3\ln c}{7}\Big ] 6^{n-m} (1+o(n)) \cr\cr
& \leq & \frac {27}{7}6^{n-m}(1+o(n)) \Big( 1+\frac {15\epsilon_5(m)}{4\sqrt{c}} \Bigr ) \sum_{j=0}^{n-m-1} \Big( \frac {\epsilon_5(m)}{8\sqrt{c}} \Big )^{2^j} \cr\cr
& \leq & \frac {27 \epsilon_5(m)6^{n-m}(1+o(n)) \Big ( 1+\frac {15\epsilon_5(m)}{4\sqrt{c}} \Big ) } {7[8\sqrt{c}-\epsilon_5(m)]} \ .
\label{Delta5upper}
\eeqs
The proof is completed by taking the infinite $n$ limit in Eq. (\ref{Ssg5}).
\ $\Box$

\bigskip

The difference between the upper and lower bounds for $S_{SG_5}$ quickly converges to zero as $m$ increases, and we have the following proposition.

\bigskip

\begin{propo} \label{proposg5} The entropy for the number of dimer coverings on the five-dimensional Sierpinski gasket $SG_5(n)$ in the large $n$ limit is $S_{SG_5}=0.67042810305...$. 

\end{propo}

\bigskip

The numerical value of $S_{SG_5}$ can be calculated with more than a hundred significant figures accurate when $m$ in Eq. (\ref{Ssg5b}) is equal to six. It is too lengthy to be included here and is available from the authors on request. 

\bigskip

We notice that the convergence of the upper and lower bounds of the entropy for dimer coverings on $SG_d(n)$ is about the same for $d=3,4,5$, similar to the results observed in \cite{dms} for the dimer-monomer model on $SG_d(n)$.

\section{Summary}
\label{sectionV}

Compare the present results with those in Ref. \cite{dms}, it is clear that the number of dimer coverings on the Sierpinski gasket $SG_d(n)$ is less than that of dimer-monomers. The asymptotic growth constant $z_{SG_{d,b}}$ for the dimer-monomer model defined as Eq. (2.1) of \cite{dms} corresponds to the entropy $S_{SG_{d,b}}$ for the dimer coverings defined in Eq. (\ref{sdef}). 
We summarize the values of $S_{SG_{d,b}}$ and the ratio $S_{SG_{d,b}}/z_{SG_{d,b}}$ in Table \ref{stable}. The value of $S_{SG_d}$ increases as dimension $d$ increases. Similarly for the generalized two-dimensional Sierpinski gasket, the exact value of $S_{SG_{2,b}}$ increases slightly as $b$ increases. For the cases studied, the ratio $S_{SG_d}/z_{SG_d}$ also increases as dimension $d$ increases, and $S_{SG_{2,b}}/z_{SG_{2,b}}$ increases slightly as $b$ increases.

It is interesting to compare entropy of dimer coverings on the Sierpinski gasket $SG_d$ with that on the $d$-dimensional hypercubic lattice ${\mathcal L}_d$ which is also $2d$-regular. The entropy of the square lattice was known to be $G/\pi$ \cite{fisher61}, where $G$ is the Catalan number, for decades, while the entropy of the simple cubic lattice was estimated to be 0.44647 \cite{nagle}. They are relatively larger than the entropies on $SG_d$ with $d=2, 3$ presented here. The values of $S_{{\mathcal L}_d}$ and the ratio $S_{SG_{d}}/S_{{\mathcal L}_d}$ for $d=2, 3$ are given in Table \ref{stable}. It appears that as the $d$ increases, the value $S_{SG_{d}}$ approaches to the value $S_{{\mathcal L}_d}$ from below. As we have obtained the highly accurate value for the entropy on $SG_d$ with $d=4, 5$, there is no numerical estimation for the entropy on ${\mathcal L}_d$ with $d \ge 4$, to the best of our knowledge. 

\begin{table}
\caption{\label{stable} Numerical values of $S_{SG_{d,b}}$, $S_{{\mathcal L}_d}$ and the ratios $S_{SG_{d,b}}/z_{SG_{d,b}}$, $S_{SG_{d}}/S_{{\mathcal L}_d}$. The last digits given are rounded off.}
\begin{center}
\begin{tabular}{|c|c|c|c|c|c|c|}
\hline\hline 
$d$ & $b$ & $D$   & $S_{SG_{d,b}}$ & $S_{SG_{d,b}}/z_{SG_{d,b}}$ & $S_{{\mathcal L}_d}$ & $S_{SG_{d}}/S_{{\mathcal L}_d}$ \\ \hline\hline 
2   & 2   & 1.585 & $\frac13 \ln 2 \simeq 0.2310490602$        & 0.3520510271 & $G/\pi \simeq 0.2915609040$ & 0.7924555624 \\ \hline
2   & 3   & 1.631 & $\frac17 \ln 6 \simeq 0.2559656385$        & 0.3811183712 & - & - \\ \hline
2   & 4   & 1.661 & $\frac{1}{12} \ln 28 \simeq 0.2776837092$  & 0.4054532859 & - & - \\ \hline
2   & 5   & 1.683 & $\frac{1}{18} \ln 200 \simeq 0.2943509648$ & - & - & - \\ \hline
3   & 2   & 2     & 0.4289638991                               &  0.5491430497 & 0.4465 & 0.9608 \\ \hline
4   & 2   & 2.322 & 0.5633747992                               & 0.6425502211 & - & - \\ \hline
5   & 2   & 2.585 & 0.6704281031                               & - & - & - \\ \hline\hline 
\end{tabular}
\end{center}
\end{table}

\bigskip

Acknowledgments: The research of S.C.C. was partially supported by the NSC
grant NSC-96-2112-M-006-001 and NSC-96-2119-M-002-001. The research of L.C.C was partially supported by TJ \& MY Foundation and NSC grant NSC 96-2115-M-030-002.

\vfill
\eject
\end{document}